\documentclass[pra,aps,footinbib,twocolumn,superscriptaddress,citeautoscript,longbibliography]{revtex4-1}

\usepackage[utf8x]{inputenc}
\usepackage{array}
\usepackage{amssymb}
\usepackage{amsmath}
\usepackage{amsfonts}
\usepackage{color}
\usepackage{graphicx}
\usepackage{bm}
\usepackage{dsfont}
\usepackage{url}
\usepackage{braket}
\usepackage{empheq}
\usepackage{tikz}
\usepackage[unicode]{hyperref}
\usepackage{color}

\hypersetup{
   unicode=true,          
   plainpages=false,
   colorlinks=true,       
   citecolor=blue,        
}
\allowdisplaybreaks
\graphicspath{{.}{./fig/}}

\newcommand{\rme}{\mathrm{e}}
\newcommand{\rmi}{\mathrm{i}}

\newcommand{\x}{\mathrm{x}}
\newcommand{\y}{\mathrm{y}}

\newcommand{\q}{{\textbf{q}}}
\newcommand{\R}{{\textbf{R}}}

\newcommand{\EFB}{\lambda}
\newcommand{\IM}{\mathbb{I}}

\newcommand{\mH}{\mathcal{H}}

\newcommand{\mhq}{\mH_q}

\newcommand{\pcsadd}{Center for Theoretical Physics of Complex Systems, Institute for Basic Science (IBS), Daejeon 34126, Republic of Korea}

\begin{document}

\title{Necessary and sufficient conditions for flat bands in $M$-dimensional $N$-band lattices with complex-valued nearest-neighbour hopping}
\author{L. A. Toikka}
\email{lauri.toikka@gmail.com}
\affiliation{\pcsadd}
\affiliation{Institute for Theoretical Physics, University of Innsbruck, A-6020 Innsbruck, Austria}
\author{A. Andreanov}
\affiliation{\pcsadd}
\date{\today}

\begin{abstract}
    We formulate the necessary and sufficient conditions for the existence of dispersionless energy eigenvalues (so-called `flat bands') and their associated compact localized eigenstates in $M$-dimensional tight-binding lattices with $N$ sites per unit cell and complex-amplitude nearest-neighbour tunneling between the lattice sites. The degrees of freedom $M$ can be traded for longer-range complex hopping in lattices with reduced dimensionality. We show the conditions explicitly for $(M = 1, N\leq4)$, $(M = 2, N = 2,3)$, and $(M = 3, N = 2,3)$, and outline their systematic construction for arbitrary $N$, $M$. If and only if the conditions are satisfied, then the system has one or more flat bands. By way of an example, we obtain new classes of flat band lattice geometries by solving the conditions for the lattice parameters in special cases. 
\end{abstract}

\maketitle

\section{Introduction}

Many interesting phases of matter exist when interactions are strong compared to the kinetic energy. Indeed, the study of strongly-interacting many-body quantum systems, ranging from high-temperature superconductors~\cite{Chen20051}, ultra-cold atoms~\cite{Zwerger2012} and quantum spin models~\cite{tsvelik2012new} to the physics of the early Universe~\cite{1367-2630-14-11-115009}, has received a lot of attention. In the case of the fractional quantum Hall effect~\cite{PhysRevLett.50.1395} (FQHE), the electrons form a strongly-correlated quantum many-body state with unusual properties such as fractionally charged excitations. The FQHE normally occurs in a two-dimensional electron gas subject to a perpendicular magnetic field, where the Landau levels have vanishing dispersion. In completely dispersionless bands, known as flat bands, the kinetic energy is formally zero, and any interaction can be considered strong. These special properties render flat band systems an excellent test bed for novel phases that appear once disorder and/or interactions are switched on: it has been shown that in principle FQHE states should be obtainable provided one can generate an approximately dispersionless band that is energetically isolated and topologically non-trivial, without the need for a magnetic field and its Landau levels~\cite{Sheng11,PhysRevLett.106.236804,PhysRevX.1.021014,doi:10.1142/S021797921330017X,PhysRevB.91.165402}. Other examples of interesting phenomena occurring in perturbed flat bands are unusual localisation properties in the presence of disorder~\cite{flach2014detangling, leykam2017localization} or quasiperiodic potential~\cite{bodyfelt2014flatbands,danieli2015flatband}, and application of an electric field leads to non-conventional Bloch oscillations~\cite{khomeriki2016landau}.

The above considerations highlight the importance of finding and identifying novel classes of flat band Hamiltonians in a systematic way. The absence of dispersion results from destructive interference. Apart from simple cases it is not clear how to engineer (nearly) dispersionless bands in general, topological or not, because this requires significant fine-tuning of the lattice overlap integrals -- flat bands are generally fragile and destroyed by perturbations unless protected by a symmetry that is respected by the perturbations.  An obvious idea of flattening a band by dividing the Hamiltonian in the momentum representation by one of the eigenstates leads to a Hamiltonian in real space that has unphyscial long-range hoppings. In contrast, much progress has been made in developing methods to discover flat band Hamiltonians with finite-range hoppings. Several approaches have been suggested to construct such flat bands: using line graphs~\cite{mielke1991ferromagnetism}, local cell construction~\cite{tasaki1992ferromagnetism}, origami rules in decorated lattices~\cite{dias2015origami} or repetition of mini-arrays~\cite{morales2016simple}. In some cases symmetries of the lattice or model can help find the flat band Hamiltonians~\cite{ramachandran2017chiral,roentgen2018compact}. The majority of these approaches, such as the line graph and the chiral constructions, however, explore properties of specific lattices. 

The eigenstates of flat band Hamiltonians with finite-range hoppings can be chosen to have strictly finite support, and correspondingly are commonly referred to as compact localised states (CLSs). Recently, much attention has been devoted to attempts at classifying and generating flat bands based on CLSs by identifying the parameter $U$, the minimum number of unit cells that a CLS associated with the flat band occupies. The complete characterisation of possible $U = 1$ flat bands has been performed in Ref.~\cite{flach2014detangling}. The case of $U = 2$ for 1-dimensional 2-band lattices with nearest-neighbour hopping is reported in Ref.~\cite{maimaiti2017compact}. This analysis can be extended to arbitrary $U$ in $d=1$~\cite{maimaiti2018unpub}. However, general progress seems problematic in higher dimensions where, for example, the number of different shapes of a given $U$ compact localised state grows rapidly with system complexity and $U$ itself.

To address the problem of finding flat bands, we provide here a general set of necessary and sufficient conditions for the occurrence of dispersionless bands in $M$-dimensional tight-binding lattices with $N$ sites per unit cell and complex-amplitude nearest-neighbor tunneling between the lattice sites. We report a new general approach for constructing flat band Hamiltonians by using the momentum representation, which has received much less attention in the flat band literature~\cite{rhim2018classification}. Our approach is distinct in that it is not restricted by the real-space property $U$ at all, while being fully general and exhaustive. We formulate a set of coupled non-linear equations for the matrix elements of the Hamiltonian and the flat band energy, and show that they form the necessary and sufficient conditions for flat bands in the system. To construct flat bands, this set of equations needs to be solved for the desired parameters. For example, the new conditions allow us to fine-tune a given Hamiltonian with a few free parameters to the nearest flat band system, or exhaustively indicate whether no flat band solutions can be found. In other words, knowing the new conditions not only answers the question of existence, but also makes it possible, in principle at least numerically, to obtain the exhaustive set of points in the space of Hamiltonians that correspond to one or more flat bands.

\section{Lattice model}

Many problems in quantum physics involve solving the Schr\"odinger equation where the potential energy is periodic in space. If for the lattice geometry a local basis in terms of on-site orbitals is sufficient, one typically models the wavefunction using the well-established tight-binding approximation. The associated Bloch Hamiltonian can be viewed as a lattice Fourier transform of the set of real-space overlap (hopping) matrices $H_{1,m}$ that describe tunneling events between the lattice sites. We write the Bloch Hamiltonian in the form
\begin{equation}
    \label{eqn:MdimBlochHam}
    \mhq = \sum_{m=1}^M\left\lbrace \rme^{\rmi \q \cdot \R _m  } H_{1,m}^\dagger +  \rme^{-\rmi \q \cdot \R _m} H_{1,m}\right\rbrace  + H_0,
\end{equation}
where the sum runs over all the unit cells that are connected through hopping, being not necessarily nearest neighbours.  Here $H_{1,m}$ (for all $m = 1, 2, \ldots, M$) and $H_0$ are $N\times N$ matrices, where $N$ is the number of bands, which equals to the number of sites in a unit cell. The matrix $H_0$ specifies the intra-cell dynamics, and the set $H_{1,m}$ specifies the inter-cell dynamics. Without loss of generality, for Hermitian systems, it suffices to consider the `canonical' case $H_0 = \mathrm{diag}(b_1,b_2,\ldots, b_N)$, where the $b_n$ (for all $n = 1, 2, \ldots, N$) are real numbers, by simply performing a suitable basis transformation. In the case of two bands ($N = 2$), one can further set $b_1=0$ and $b_2=1$ by suitable shifts and rescaling of the spectrum of the Hamiltonian. Quite generally, hopping amplitudes in a lattice can have a phase, e.g. from the Aharonov-Bohm effect in the presence of a magnetic field or spin-orbit coupling; we therefore take $H_{1,m} \in \mathbb{C}$.
 
In this work, we require the flat band to be dispersionless with respect to the $M$ separate momenta $q_m =  \q \cdot \R _m$. Importantly, we do not specify the momenta, which means that they can be taken to be linearly independent or linearly dependent. The former case describes an $M$-dimensional lattice with strictly nearest-neighbour hopping, while in the latter case a subset of the momenta $q_m$ has been utilized to describe lattices that include further nearest-neighbour hopping. In particular, this flexibility makes our approach powerful for describing a plethora of different flat band lattice geometry classes once we have solved for $H_{1,m}$  and $H_0$ the conditions that guarantee the existence of one or more flat bands.

\section{$N$ bands in one dimension with complex hopping}

To determine the set of necessary and sufficient conditions for one or more flat bands, we start by writing the characteristic equation with eigenvalue $\EFB$,
\begin{equation}
    \det{\left(\mhq - \EFB\IM\right)} = 0, \qquad \forall q,
\end{equation}
which, as indicated, is intended to hold for all $q$ guaranteeing the non-dependence of $\EFB$ on momentum in the entire Brillouin zone.  Focusing in this Section on nearest-neighbour hoppings, $H_1\equiv H_{1,m}$, the Bloch Hamiltonian reads
\begin{equation}
    \mhq = \rme^{\rmi q} H_1^\dagger + H_0 +  \rme^{-\rmi q} H_1,
\end{equation}
where $H_1$ and $H_0$ are $N\times N$ matrices, and $N$ is the number of bands.

To find the necessary and sufficient flat band conditions, we first express the characteristic polynomial $p(\EFB) \equiv \det{(\mhq - \EFB \IM_N)} = \sum_{k=0}^N c_k \EFB^k$ of the $N\times N$ matrix $\mhq$ as a polynomial in $\eta \equiv \rme^{\rmi q}$: $p(\eta) = \sum_{k = -N}^N \tilde{x}_k \eta^k$, where $\tilde{x}_k \equiv \tilde{x}_{-k}^*$. We then solve the coupled non-linear set $\{ \tilde{x}_k\} = 0$ for the matrix elements of $H_1$ and $H_0$. It can be proven by induction that this set of equations forms the necessary and sufficient existence conditions for flat bands in one dimension. We generalise this principle to higher dimensions in Sec.~\ref{sec:higherD}.

\subsection{Example: two bands in one dimension}

The procedure described above assumes its simplest non-trivial form in the case of two bands in one dimension with real-valued nearest neighbour hoppings, for which the Bloch Hamiltonian reads
\begin{equation}
    \mhq = \rme^{\rmi q} H_1^\mathrm{T} + H_0 +  \rme^{-\rmi q} H_1.
\end{equation}
In this section, $H_1 =  \begin{pmatrix}
a & b\\
c & d\\
\end{pmatrix}$ with $(a,b,c,d)  \in \mathbb{R}^4$, and we assume that $H_0$ is scaled so that it is given by $\begin{pmatrix}
0 & 0\\
0 & \Theta(\zeta)\\
\end{pmatrix}$, where $\Theta$ is the Heaviside step function and $\zeta = \pm 1$ is a parameter. Then
\begin{equation}
    \label{eqn:LPFBgen-1a}
    \begin{split}
        \det{\left(\mhq - \EFB \sigma_0 \right)} 
        &\equiv \left(\eta ^2+\frac{1}{\eta ^2}\right) x_2+\left(\eta +\frac{1}{\eta }\right) x_1 +x_0 ,
    \end{split}
\end{equation}
where $\eta = \rme^{\rmi q}$, and
\begin{subequations}
    \label{eqn:LPFBgen-2}
    \begin{align}
        x_0 &=  2 a d-b^2-c^2 -\EFB \Theta (\zeta )+\EFB^2,\\
        x_1 &= a  \Theta(\zeta) - \EFB(a+d),\\
        x_2 &= a d - b c.
    \end{align}
\end{subequations}
Setting $(x_0,x_1,x_2) = (0,0,0)$, we find from the system~\eqref{eqn:LPFBgen-2} the necessary and sufficient flat band conditions
\begin{subequations}
    \label{eqn:LPFBgen-2bands}
    \begin{align}
        -\EFB \Theta (\zeta ) + \EFB^2  &= (b-c)^2,\\
        b c &= a d,\\
        -\EFB  (a+d) &= a  \Theta(\zeta),
    \end{align}
\end{subequations}
which can be straightforwardly solved for $(a,b,c,d)$ once $\zeta$ and $\EFB$ are given. For example, a flat band at zero energy ($\EFB = 0$) exists such that
\begin{itemize}
    \item If $\zeta = 1$ (on-site terms), then $(a,b,c,d) = (0,0,0,d)$ is the only solution with $d$ arbitrary.
    \item If $\zeta = -1$ (no on-site terms), then there are infinitely many solutions such that $b = c$, $ad = bc$.
\end{itemize}

In the context of the classifications of flat bands based on the compact localised states, both of the above cases have the property $U=1$, since $\EFB$ coincides with an eigenvalue of $H_0$~\cite{maimaiti2017compact}. The algebraic equations~\eqref{eqn:LPFBgen-2bands} can be solved for the Hamiltonian with general $\EFB$ and $\zeta$, in which case our approach independently reproduces the results and exactly reduces to the flat band generator of Ref.~\cite{maimaiti2017compact}. It is instructive to note that given a flat band eigenvector in the Bloch representation, it is straightfoward to extract the CLS, although it is not guaranteed to be the smallest possible CLS for that flat band.

\subsection{Determinant expansion for the necessary and sufficient flat band conditions}

Since the determinant of a Hermitian matrix is always real, in the expansion of the determinant the complex-valued terms $\eta^k\tilde{x}_k$ are always be paired with their complex conjugates. The polynomial $p(\eta)$ reads 
\begin{equation}
    \label{eqn:LPFBgen-9-complex}
    \begin{split}
        & \det{\left(\mhq - \EFB \IM_N \right)} =  \\
        & \;\; \frac{\tilde{x}_{-1}}{2}   + \frac{1}{\eta}\tilde{x}_{-2} + \frac{1}{\eta^2} \tilde{x}_{-3} + \ldots + \frac{1}{\eta^N} \tilde{x}_{-(N+1)}  + \mathrm{c.c.},
    \end{split}
\end{equation}
where $\tilde{x}_i = \tilde{x}_{-i}^*$ for all $i = 1, 2, \ldots, N+1$. Here, the coefficients $\tilde{x}_i$ are in general non-linear functions of the hopping and on-site parameters determining $H_0$ and $H_1$.  As indicated earlier, to formulate the flat band conditions we must have all the coefficients vanish individually: $\tilde{x}_{-1} = \tilde{x}_{-2}  = \ldots = \tilde{x}_{-(N+1)} = 0$. 

After a direct calculation, we find
\begin{subequations}
    \label{eqn:LPFBgen-Nbands-syst-complex}
    \begin{align}
    \label{eqn:LPFBgen-Nbands-syst-1-complex-2}
        \tilde{x}_{-(N+1)} = 0 &=  \det\,H_1,\\
    \label{eqn:LPFBgen-Nbands-syst-2-complex-2}
        \tilde{x}_{-N} = 0 &= \sum_{k = 1}^N\left[ \left(H_0\right)_{kk}  - \EFB \right] C_{kk} ,\\
    \label{eqn:LPFBgen-Nbands-syst-3-complex-2}
        \tilde{x}_{-(N-1)} = 0 &=  \mathrm{tr} \left( H_1^*  C\right)+ \mathrm{tr}{\left(H_1 C_0 \right)} \\ 
        \notag & \qquad + \tilde{f}_{-(N-1)}^{(N)}(\EFB, H_0,H_1),\\
        \notag &\ldots\\
    \label{eqn:LPFBgen-Nbands-syst-4-complex}
        \tilde{x}_{-1} = 0 &= \tilde{f}_{-1}^{(N)}(\EFB, H_0,H_1) \\
        & \mathrm{+ c.c.}
    \end{align}
\end{subequations}
Here $C$ ($C_0$) is the cofactor matrix of $H_1$ ($H_0$), and $H_0$ is taken to be diagonal and real. Specifically, we define $C_{ij} = (-1)^{i+j}M_{ij}$ to be the $(i,j)$th cofactor of $H_1$, where $M_{ij}$ is the $(i,j)$th minor of $H_1$. The functions $\tilde{f}$ are shown explicitly for $N \leq 4$ with $H_1 \in \mathbb{C}$ in system~\eqref{eqn:LPFBgen-Nbands-syst-addprop1-complex} in Appendix~\ref{app:B}.

The system~\eqref{eqn:LPFBgen-Nbands-syst-complex} amounts to the full set of sufficient and necessary existence conditions for a flat band at energy $\EFB$ and its associated compact localized eigenstates in one dimension with $N$ bands and complex nearest-neighbour hopping. That they are sufficient is trivial, and that they are also necessary can be proven by induction. We identify the condition~\eqref{eqn:LPFBgen-Nbands-syst-1-complex-2} as the self-interference property of CLSs that requires the hopping matrix $H_1$ to be degenerate. It reflects the physical property that the flat band eigenstates are compact, that is, there exists a non-zero eigenvector $\psi$ such that $H_1 \psi = 0$ which means at least one of the eigenvalues of $H_1$ must be zero leading to Eq.~\eqref{eqn:LPFBgen-Nbands-syst-1-complex-2}. However, as we have shown here explicitly in terms of the exhaustive existence conditions~\eqref{eqn:LPFBgen-Nbands-syst-complex}, one evidently needs much more to suppress dispersion. We have verified Eqs.~\eqref{eqn:LPFBgen-Nbands-syst-1-complex-2},~\eqref{eqn:LPFBgen-Nbands-syst-2-complex-2}, and~\eqref{eqn:LPFBgen-Nbands-syst-3-complex-2} with the supplementary property $\tilde{f}_{-(N-1)}^{(N)}(0,0,H_1) = 0$ for $N \leq 8$.

Mathematically, Eq.~\eqref{eqn:LPFBgen-9-complex} constitutes a determinant expansion for the $N\times N$ matrix $\eta H_1^\dagger + H_0 + \eta^* H_1 - \EFB\IM_N $ as a power series in $\eta$. While we have computed it directly for $N \leq 5$, the algorithmic procedure for constructing the functions $\tilde{f}_{-(N-i)}^{(N)}$ for general $(N,i = 1,2,\ldots N-1)$ is limited only by the capacity of generic symbolic computation packages. The principle of the above proof by construction for the necessary and sufficient CLS-FB conditions is the same for any $N$.

\section{Higher dimensions, complex hopping}
\label{sec:higherD}

Having an $M$-dimensional Bloch Hamiltonian~\eqref{eqn:MdimBlochHam} means that the conditions~\eqref{eqn:LPFBgen-Nbands-syst-complex} for the one-dimensional case are repeated for every $m = 1, \ldots, M$. However, there are also additional conditions from the cross-term coefficients. The multinomial $p(\{\eta_m \})$ reads
\begin{equation}
    \label{eqn:LPFBgen-9-complex-Mdim}
    \begin{split}
        \det{\left(\mhq - \EFB\IM_N \right)}
        &= \left\lbrace \sum_{m = 1}^M  \sum_{n = -N}^{-1} \eta_m^n \left( \tilde{x}_n^{(m)} + \hat{x}_n^{(m,N)}  \right)  + \mathrm{c.c.} \right\rbrace \\
        & + F_0^{(N,M)} + \sum_{\{ k_m \} } \left( \prod_{m=2}^M \eta_m^{k_m}\right) y_{\{ k_m \} }^{(N,M)} ,
    \end{split}
\end{equation}
where $m$ is the dimensionality index, $n$ is the band number index, $\eta_m \equiv  \rme^{\rmi \q \cdot \R_m  }$, and the set $\{ k_m \} \equiv k_1,k_2,\ldots,k_M$. We have $x_{n}^{(m)} = \bar{x}_{-n}^{(m)}$, and $y_{\{ k_m \} }^{(N,M)}  = \bar{y}_{-\{ k_m \} }^{(N,M)}$ because the determinant is real. Note: $y_{\{ k_m \} }^{(N,M)}  = 0$ if $\{ k_m \}$ has only one non-zero element. Here $x_n^{(m)} = \tilde{x}_n^{(m)} + \hat{x}_n^{(m,N)}  $, where $ \hat{x}_n^{(m,N)} $ is the additional contribution to the direct one-dimensional coefficient. The algebraic form~\eqref{eqn:LPFBgen-9-complex-Mdim} exhausts every possibility; we specify the only non-zero cross-terms $y$ together with the $\hat{x}$ and  $F_0^{(N,M)}$ terms for most common cases of interest in Appendix~\ref{app:A}. 

The necessary and sufficient CLS-FB existency equations for $N$ bands in $M$ dimensions form the set of equations $F_0^{(N,M)} = 0 \cap \{x_{n}^{(m)}  = 0\} \cap \{y_{\{ k_m \} }^{(N,M)}  = 0\}$ for all $m= 1,2, \ldots, M$ and $n = -N, -N + 1, \ldots, -1$. For all $m$:
\begin{subequations}
    \label{eqn:LPFBgen-Nbands-syst-complex-Mdim}
    \begin{align}
    \label{eqn:LPFBgen-Nbands-syst-1-complex-2-Mdim}
        0 &=  \det{(H_{1,m})} + \hat{x}_{-N}^{(m,N)} , \\
    \label{eqn:LPFBgen-Nbands-syst-2-complex-2-Mdim}
        0 &= \sum_{k = 1}^N\left[ \left(H_0\right)_{kk}  - \EFB \right] C_{kk}^{(m)} + \hat{x}_{-N-1}^{(m,N)} ,\\
    \label{eqn:LPFBgen-Nbands-syst-3-complex-2-Mdim}
        0 &=  \mathrm{tr} \left( \bar{H}_{1,m}  C^{(m)}\right)+ \mathrm{tr}{\left(H_{1,m} C_0 \right)}  \\
        \notag & \qquad \tilde{f}_{-(N-1)}^{(N)}(\EFB, H_0,H_{1,m}) + \hat{x}_{-N-2}^{(m,N)} ,\\
        \notag &\ldots\\
    \label{eqn:LPFBgen-Nbands-syst-4-complex-Mdim}
        0 &= F_0^{(N,M)},\\
        & \mathrm{+ c.c.}\\
        0 &= y_{\{ k_m \} }^{(N,M)} \qquad \forall \{ k_m \} ,
    \end{align}
\end{subequations}
where the bar denotes complex conjugation. The complex conjugates in effect are automatically satisfied if the above are satisfied. The functions $\tilde{f}$ are given in system~\eqref{eqn:LPFBgen-Nbands-syst-addprop1-complex} up to $N\leq 4$.

If the necessary and sufficient conditions~\eqref{eqn:LPFBgen-Nbands-syst-complex-Mdim} are satisfied, then the lattice has an energy eigenvalue $\EFB$ (there can also be several simultaneous solutions for $\EFB$) that is independent of $q_1, q_2, \ldots, q_M$. However, the individual momenta $q_m$ (i.e. the lattice vectors $\R_m$) are free; in particular, they can be linearly independent or linearly dependent. This means that a solution for the parameters $(N, M)$ can be used to describe flat bands in $N$-band systems with any $M$ kinds of hopping; e.g. nearest and next-nearest-neighbour hopping together with nearest neighbours in directions that are linear combinations of the primitive vectors of the Bravais lattice. Provided $M>1$, we can trade dimensionality of the system for longer-range hopping. 

Moreover, our approach makes it possible to construct a pseudo-flat band that is flat only along the directions $\{m_i \}$ in momentum space, by solving only the conditions involving $m \in \{m_i \}$, and leaving the directions $m  \notin \{m_i \}$ unconstrained.

\section{New classes of flat band lattice geometry}

We now demonstrate the usefulness of our apporach and solve the necessary and sufficient conditions~\eqref{eqn:LPFBgen-Nbands-syst-complex-Mdim} for example Bloch Hamiltonians where we fix enough parameters to make the conditions easily solvable for the remaining ones. This can be used to straightforwardly generate numerous flat band solutions for the parameters of the Bloch Hamiltonian. We will report the results of a systematic and exhaustive search for new flat band lattices later; here, we illustrate our approach with a selection of Hamiltonians and different classes $U$ of CLS, i.e. their sizes.

\subsection{$U = 1$ CLS}
\label{sec:U1}

If the flat band energy $\EFB$ coincides with any of the diagonal elements $b_n$ of $H_0$, then the resulting compact localized states associated with the flat band are all contained within just one unit cell~\cite{maimaiti2017compact}; these are so-called $U = 1$ flat bands~\cite{flach2014detangling} where $U$ is the number of lattice unit cells the eigenstate occupies. In this Section, we take the flat band energy $\EFB = 0$ to coincide with the diagonal elements of $H_0$ making these flat bands all belong to the class $U = 1$. We present more cases in Appendix~\ref{app:C}.

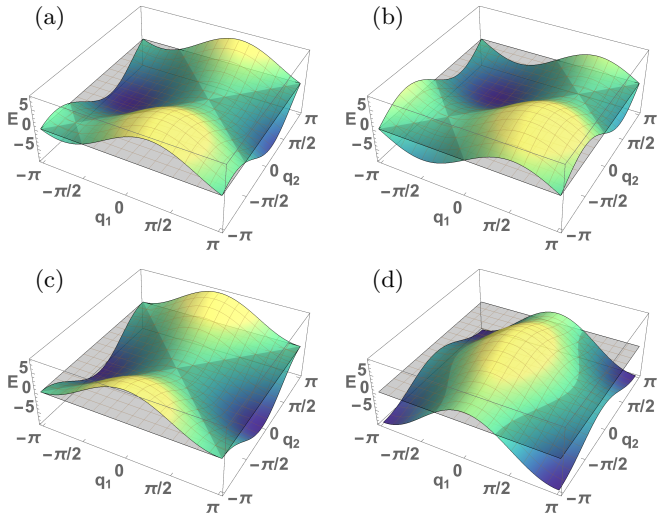
\begin{figure}[t]
  \centering
        \begin{tikzpicture}
        \def\x{4.5};        \def\y{3.5};
              \node at (0,0) {    \includegraphics[width=0.23\textwidth,angle=0]{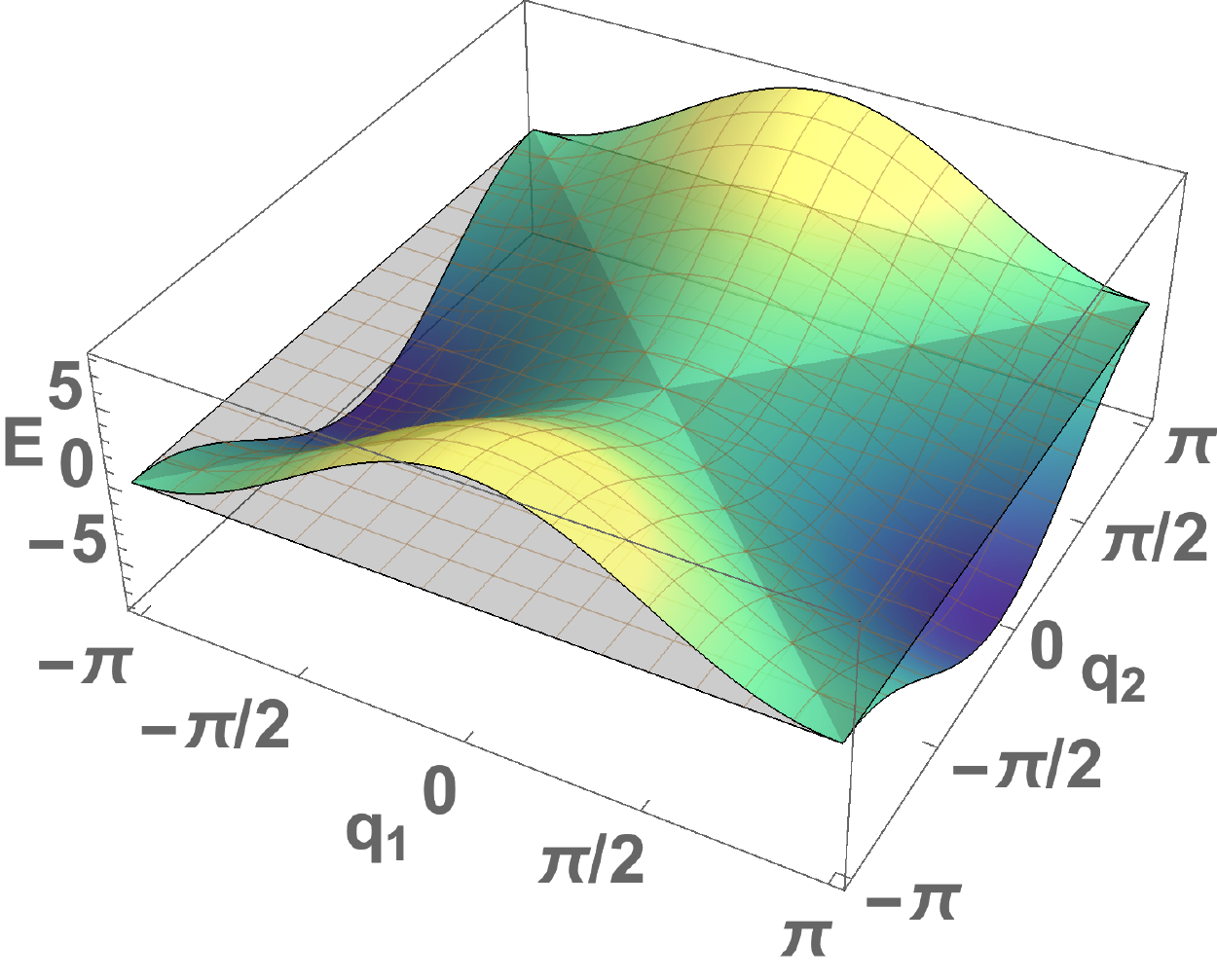}};
              \node at (\x,0) {    \includegraphics[width=0.23\textwidth,angle=0]{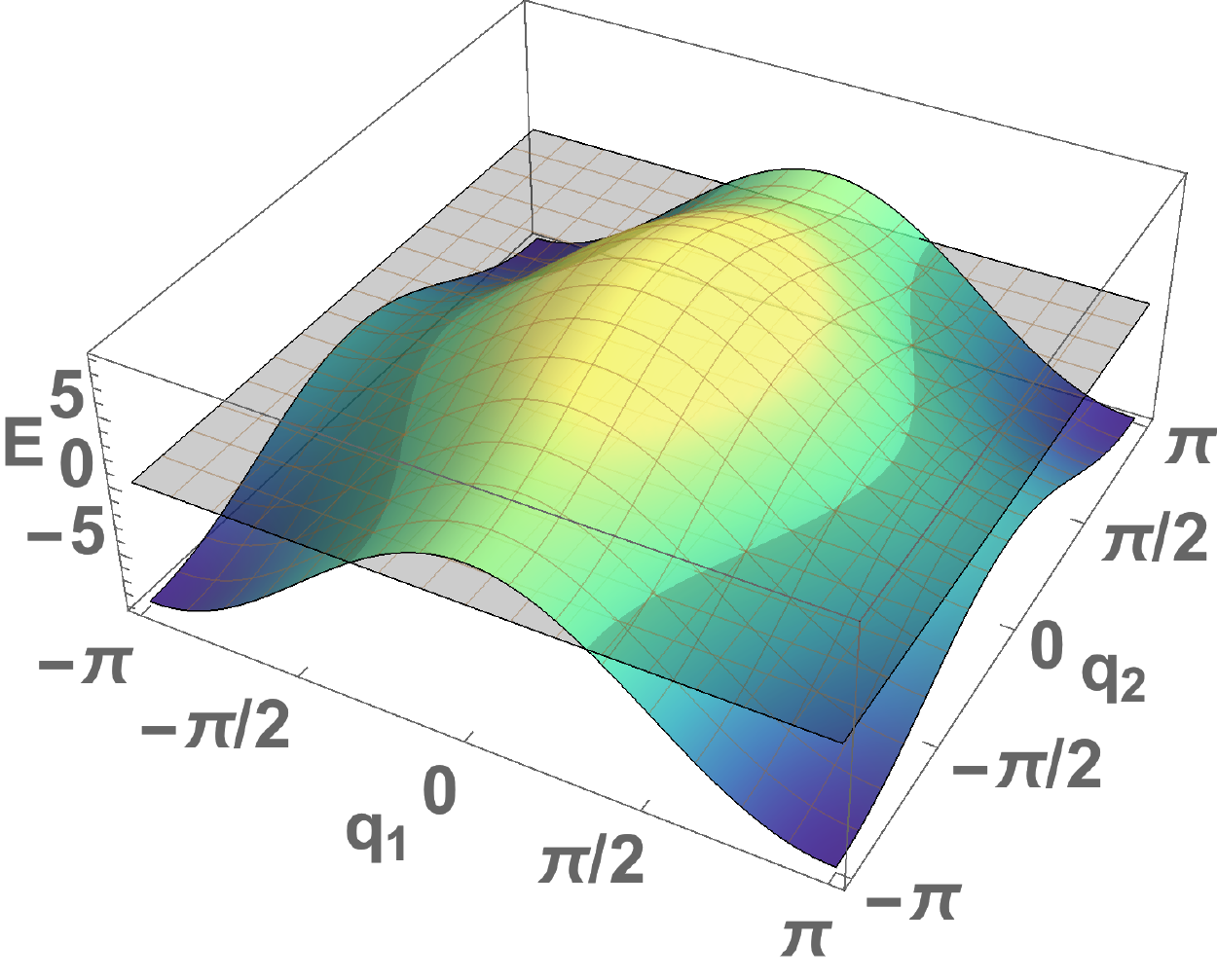}};
			 \node at (0,\y) {    \includegraphics[width=0.23\textwidth,angle=0]{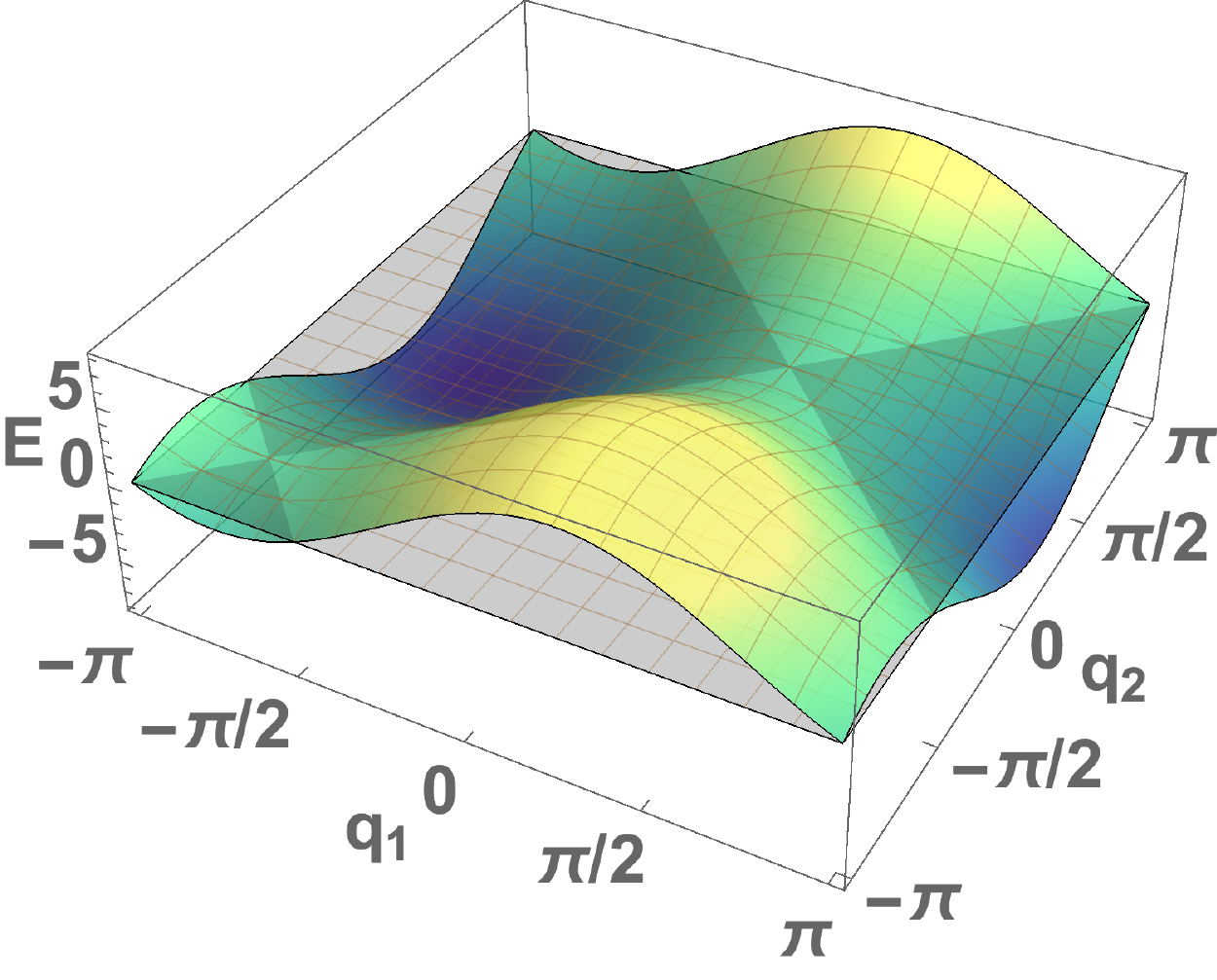}};
              \node at (\x,\y) {    \includegraphics[width=0.23\textwidth,angle=0]{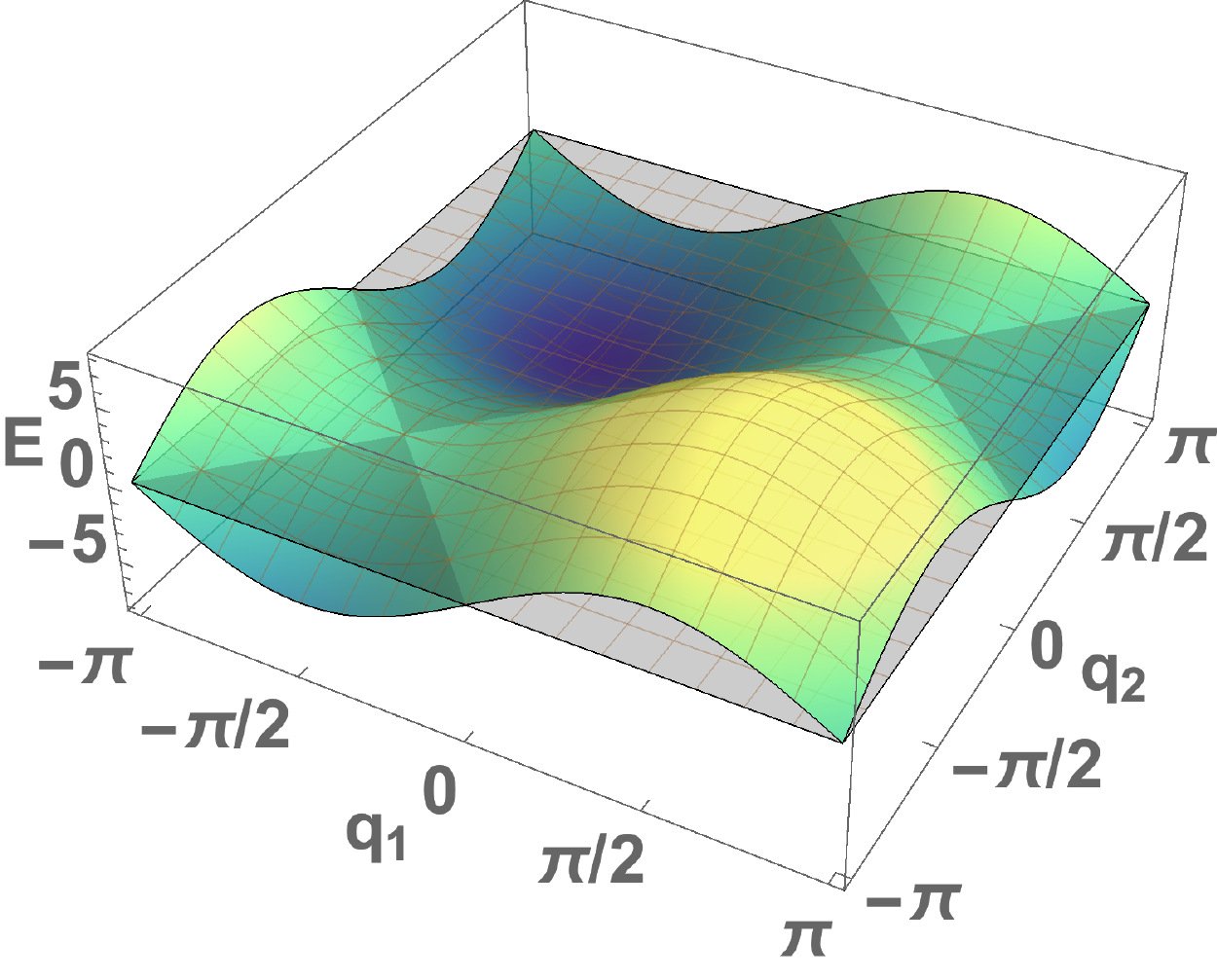}};
      \node at (-1.5,5) {(a)};
            \node at (-1.5+\x,5) {(b)};
                  \node at (-1.5,5-\y) {(c)};
                        \node at (-1.5+\x,5-\y) {(d)};
      \end{tikzpicture}
    \caption{\textit{$(2, 2)$-Flat band lattices}. Eigenvalues of the Hamiltonian~\eqref{eqn:ex22} with (a) $c_1 = \rme^{-\rmi \frac{\pi}{2} } $, $c_2 = \rme^{\rmi \frac{\pi}{4} } $; (b) $c_1 = \rme^{-\rmi \frac{\pi}{2} } $, $c_2 = \rme^{\rmi \frac{\pi}{2} } $;  (c) $c_1 = 1 $, $c_2 =1 $; and (d) $c_1 =1+\rmi$, $c_2 = 1$. The parameter $d = -1$ in (a-c) and $d = 0.5$ in (d). The band crossings reflect the general property of the $U = 1$ class that the flat band system can be detangled~\cite{flach2014detangling}, i.e. unitary transformed into disjoint systems.}
    \label{fig:1}
\end{figure} 

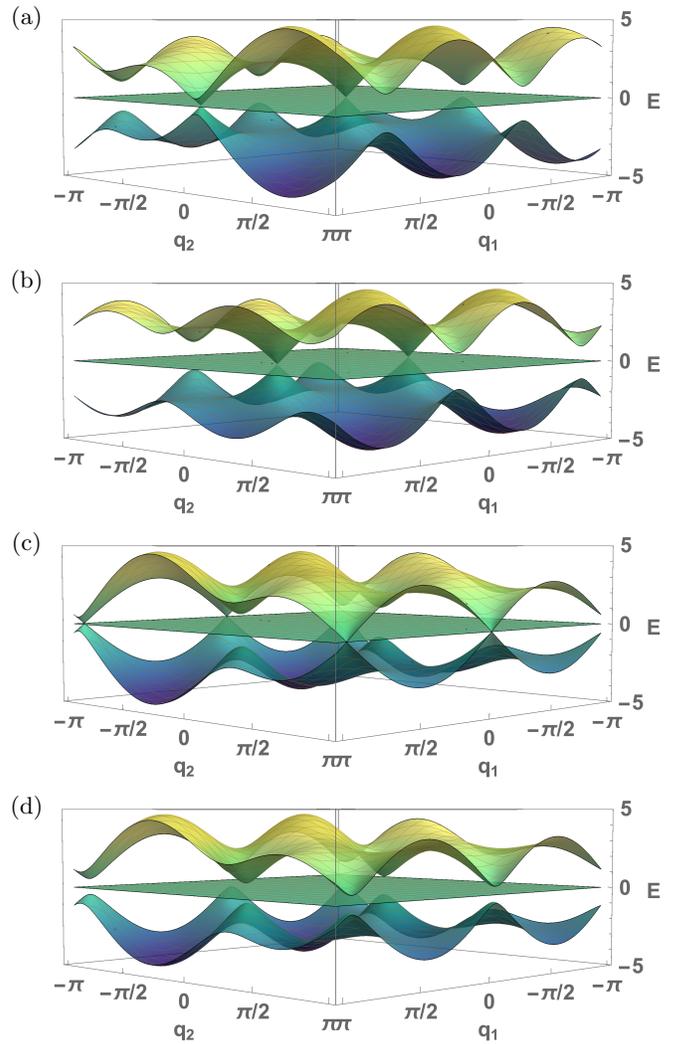
\begin{figure}[t]
  \centering
        \begin{tikzpicture}
        \def\x{-4.4};        \def\y{3.5};
              \node at (0,0) {    \includegraphics[width=0.45\textwidth,angle=0]{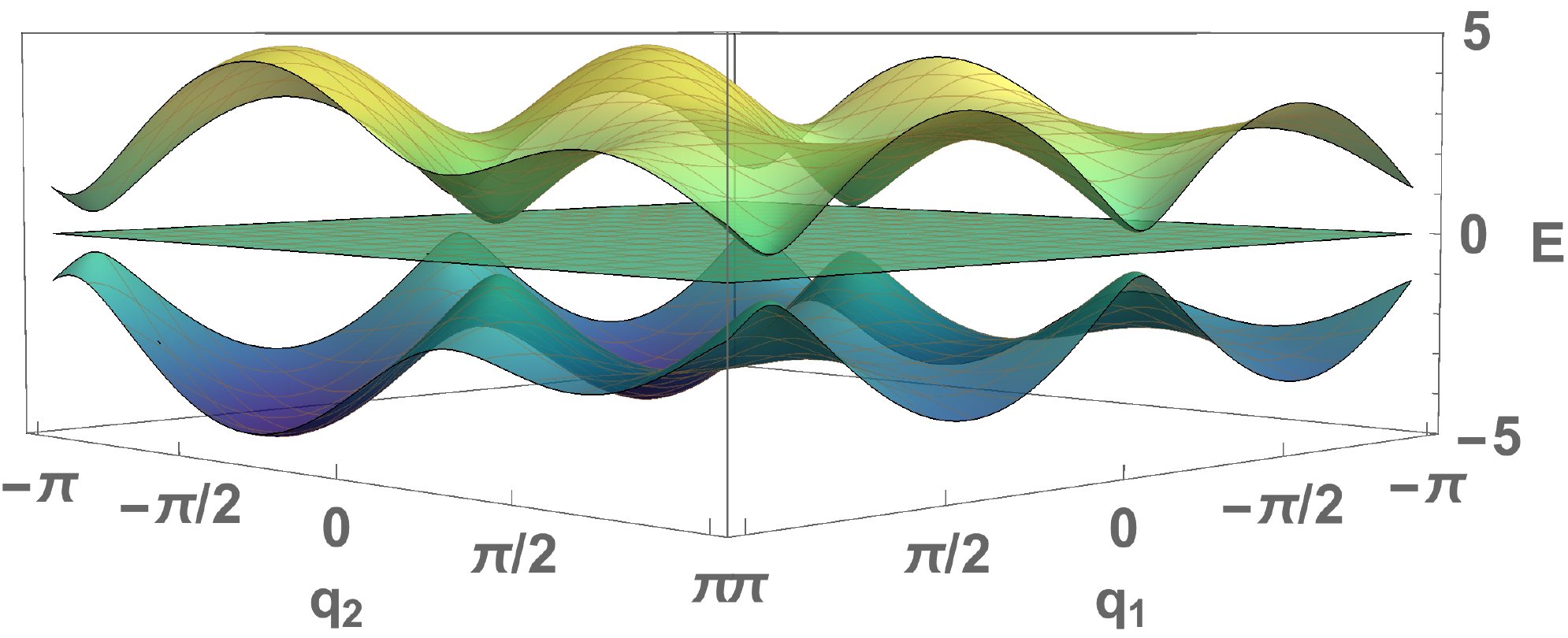}};
              \node at (0,\y) {    \includegraphics[width=0.45\textwidth,angle=0]{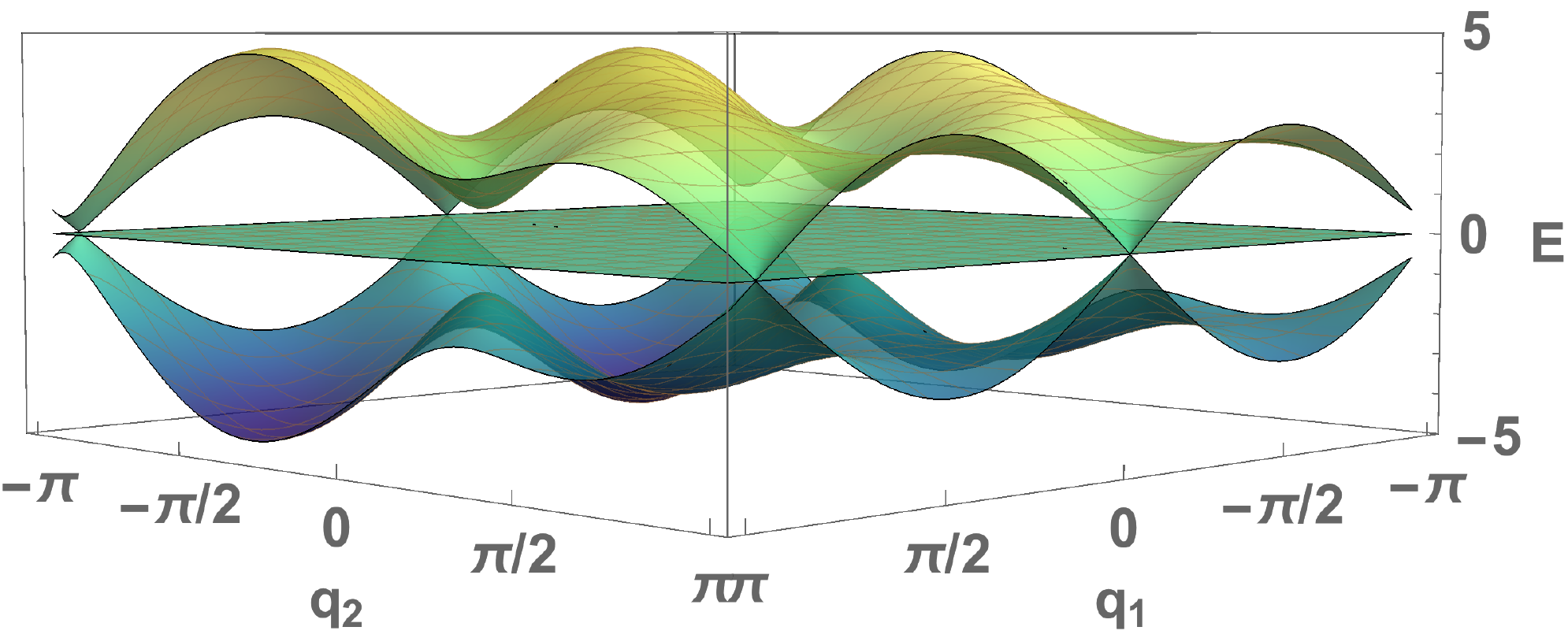}};
			 \node at (0,2*\y) {    \includegraphics[width=0.45\textwidth,angle=0]{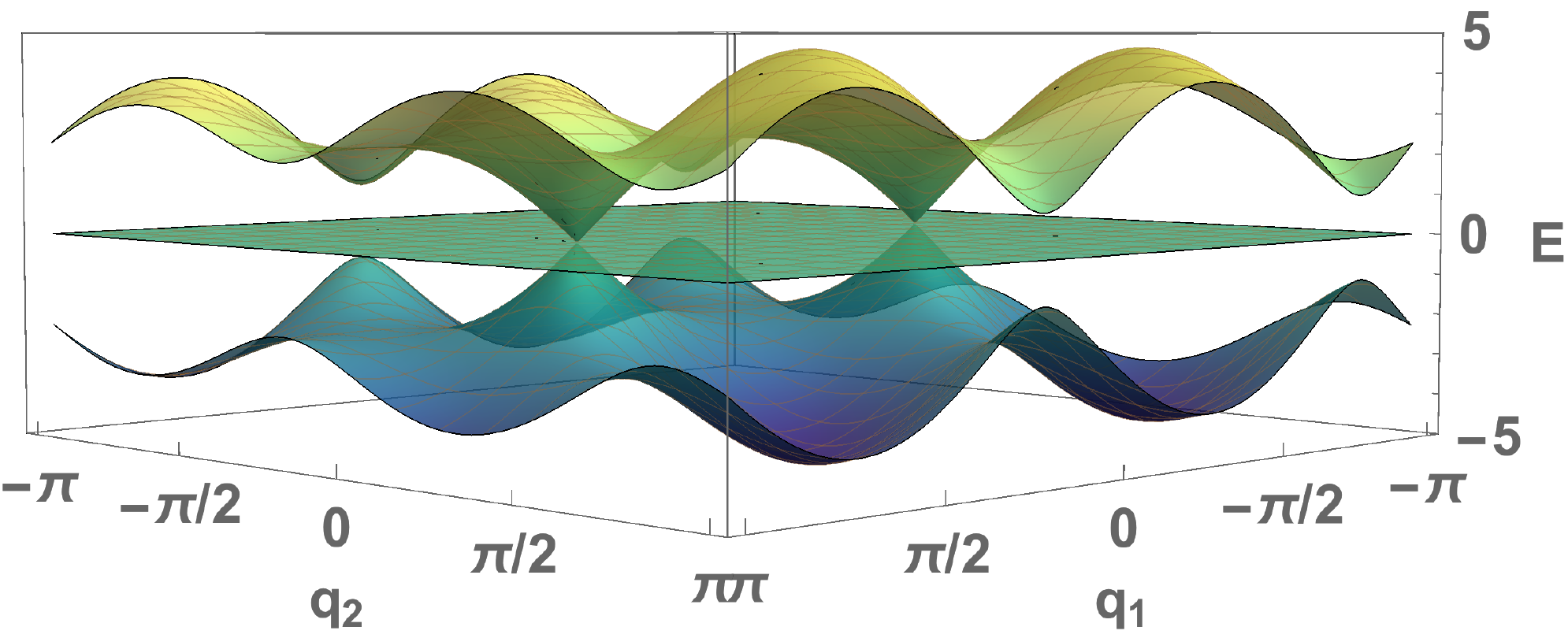}};
              \node at (0,3*\y) {    \includegraphics[width=0.45\textwidth,angle=0]{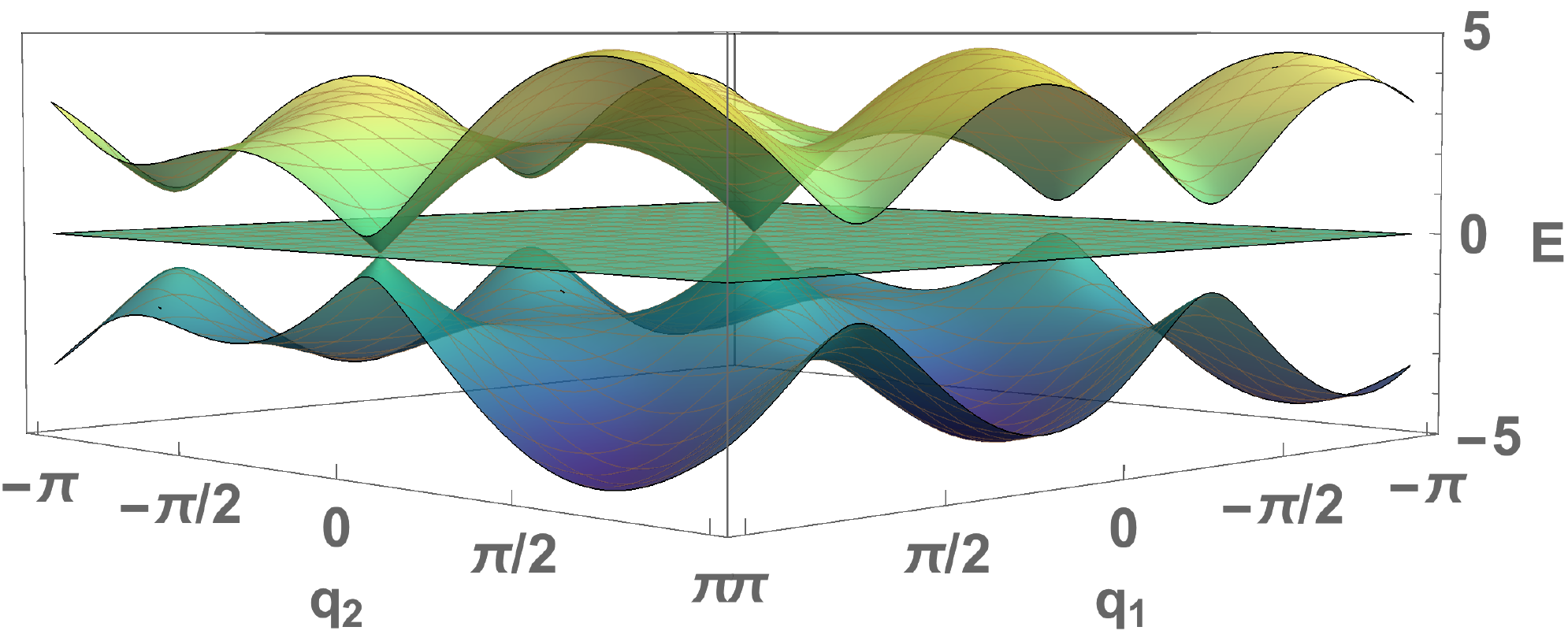}};
      \node at (\x,12) {(a)};
            \node at (\x,12-\y) {(b)};
                  \node at (\x,12-2*\y) {(c)};
                        \node at (\x,12-3*\y) {(d)};
      \end{tikzpicture}    
  
    \caption{\textit{$(3, 3)$-Flat band lattices with Weyl points}. Eigenvalues of the Hamiltonian~\eqref{eqn:ex33a} with $q_3 = q_1 - q_2$ and (a) $c_1 = -1$, $c_2 = \rme^{\rmi \frac{\pi}{2} } $, $c_3 = \rme^{\rmi \frac{\pi}{3} } $, $c_4 = \rme^{\rmi \frac{\pi}{5} } $; (b) $c_1 = 1$ and $c_3 =  \rme^{\rmi \frac{4\pi}{3} }$ otherwise same as (a); (c) $c_1 = 1$ otherwise same as (a);  (d) $c_1 =  \rme^{-\rmi \frac{\pi}{7} } $ otherwise same as (a). There are two band-touching points either inside (a) or at the boundary (b) of the Brillouin zone. In (c) there are no such points, and the flat band is separated from the two dispersive bands.}
    \label{fig:2}
\end{figure} 

\subsubsection{$N = 2$, $M = 2$}
Setting $H_0 = 0$ and $\EFB = 0$, we can solve the coupled non-linear conditions~\eqref{eqn:LPFBgen-Nbands-syst-complex-Mdim} for the remaining parameters of the matrix $H_1$ algebraically to find, among other solutions, e.g.
\begin{equation}
    \label{eqn:ex22}
    \mhq = \left[ \rme^{-\rmi q_1} + d \rme^{-\rmi q_2} \right]
    \begin{pmatrix}
        \frac{ \left|  c_1 \right| ^2}{\bar{c}_2} & c_1 \\
        \frac{c_2 \bar{c}_1}{\bar{c}_2} & c_2 \\
    \end{pmatrix} + \mathrm{h.c.},
\end{equation}
such that $d \in \mathbb{R}$, and $c_1, c_2 \in \mathbb{C}$ (Fig.~\ref{fig:1}). 
If $d  = 0$, then we have a quasi-flat band that does not depend on $q_2$. The eigenvalues of the $2\times 2$ Hamiltonian~\eqref{eqn:ex22} can be solved analytically; we find
\begin{equation}
E(q_1,q_2) =  \begin{cases}\!
     0,  &  \\
  \begin{aligned}[b]
   2 \left(1+ \frac{\left| c_1 \right| ^2}{\left| c_2 \right| ^2} \right) \bigg\lbrace  \mathrm{Re}(c_2 ) \left[\cos (q_1 )+ d\cos (q_2 )\right]  \\ 
     + \mathrm{Im} (c_2)  \left[\sin (q_1 )+ d\sin (q_2 )\right] \bigg\rbrace.
  \end{aligned} &
\end{cases}
\end{equation}

\subsubsection{$N = 3$, $M = 3$}
Similarly, setting a subset of the parameters of the Hamiltonian to zero, we find that the following are examples of flat-band solutions to the conditions~\eqref{eqn:LPFBgen-Nbands-syst-complex-Mdim}: 

\begin{subequations}
    \label{eqn:ex33}
    \begin{align}
        \label{eqn:ex33a}
        \mhq &=   \rme^{-\rmi q_1}
        \begin{pmatrix}
            0 & c_1 & 0 \\
            0 & 0 & c_1 \\
            0 & 0 & 0 \\
        \end{pmatrix} 
        + \rme^{-\rmi q_2}
        \begin{pmatrix}
            -c_2 & 0 & 0 \\
            0 & 0 & 0 \\
            0 & 0 & c_2 \\
        \end{pmatrix}\\ 
        \notag & + \rme^{-\rmi q_3}
        \begin{pmatrix}
            0 & c_3 & 0 \\
            c_4 & 0 & c_3 \\
            0 & c_4 & 0 \\
        \end{pmatrix}
        + \mathrm{h.c.}, \\
        \label{eqn:ex33b}
        \mhq &=   \rme^{-\rmi q_1}
        \begin{pmatrix}
            0 & c_1 & 0 \\
            0 & 0 & c_1 \\
            0 & 0 & 0 \\
        \end{pmatrix} 
        + \rme^{-\rmi q_2}
        \begin{pmatrix}
            0 & 0 & 0 \\
            0 & c_2 & 0 \\
            0 & 0 & 0 \\
        \end{pmatrix}\\ 
        \notag & + \rme^{-\rmi q_3}
        \begin{pmatrix}
            0 & c_3 & 0 \\
            -\frac{c_5 \bar{c}_3}{\bar{c}_4} & 0 & c_4 \\
            0 & c_5 & 0 \\
        \end{pmatrix}
        + \mathrm{h.c.},
    \end{align}
\end{subequations}
such that $c_{1-5} \in \mathbb{C}$ (Fig.~\ref{fig:2}). 

\begin{figure}[t]
  \centering

        \begin{tikzpicture}
        \def\x{4.5};        \def\y{3.5};
              \node at (0,0) {    \includegraphics[width=0.23\textwidth,angle=0]{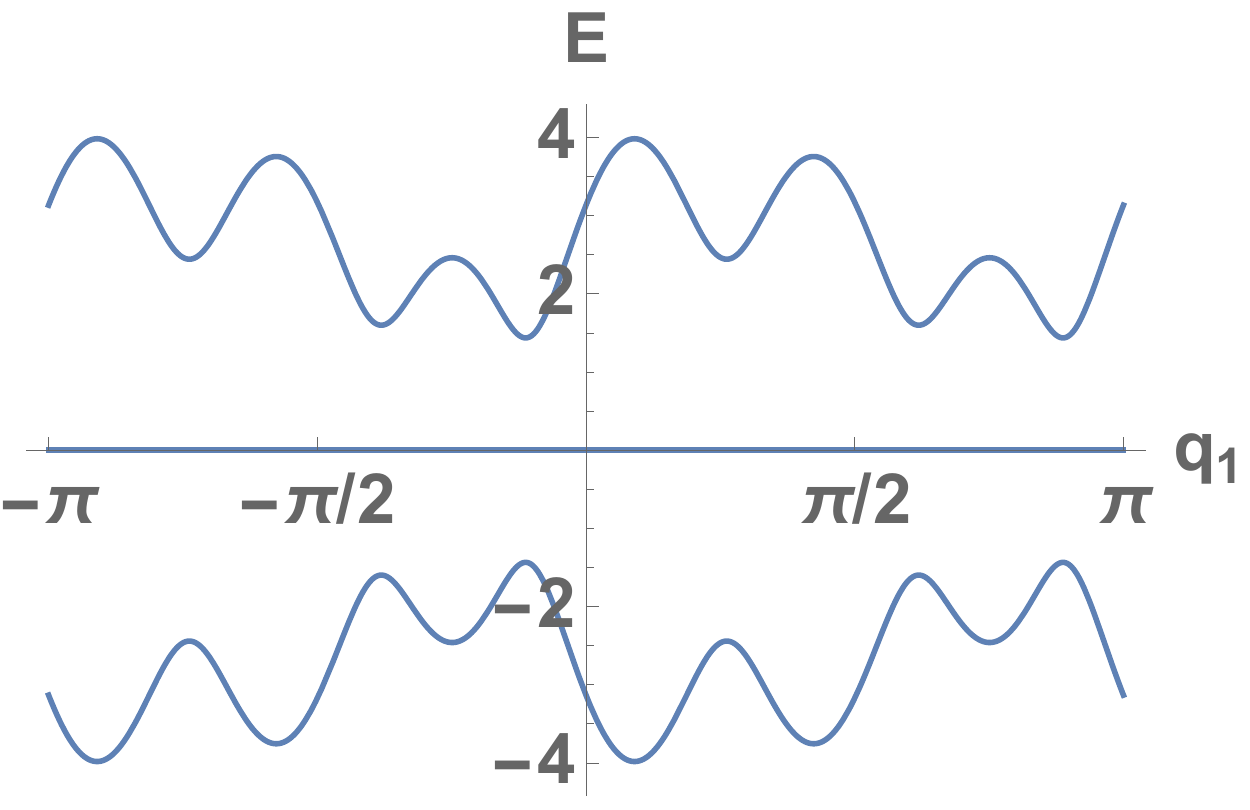}};
              \node at (\x,0) {    \includegraphics[width=0.23\textwidth,angle=0]{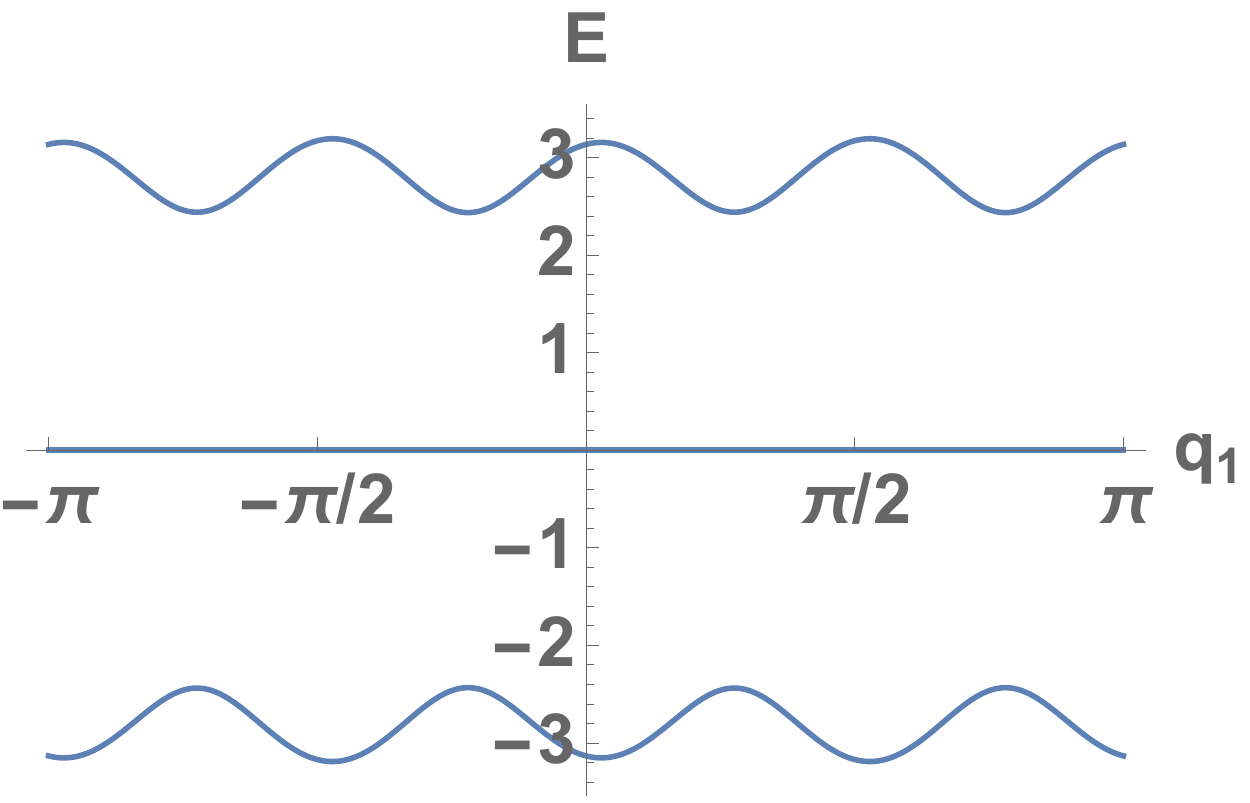}};
	    \node at (-1.5,1.3) {(a)};
            \node at (-1.5+\x,1.3) {(b)};
      \end{tikzpicture}

    \caption{\textit{Reduction of a $(3, 3)$-flat band lattice}. Eigenvalues of the Hamiltonian~\eqref{eqn:ex33a} with $c_1 = \rme^{-\rmi \frac{\pi}{7} }$, $c_2 = \rme^{\rmi \frac{\pi}{2} } $, $c_3 = \rme^{\rmi \frac{\pi}{3} } $, $c_4 = \rme^{\rmi \frac{\pi}{5} }$. In (a) $q_3 = 3 q_1$, $q_2 = 2 q_1$; (b) $q_3 = q_1$, $q_2 = 2 q_1$, and $q_1 \to 3 q_1$. Both cases correspond to nearest and next-nearest neighbour hopping.
      }
    \label{fig:3}
\end{figure}

Here, $q_3$ ($q_2$) can be any linear combination of $q_1,q_2$ ($q_1,q_3$). For example,
\begin{subequations}
    \label{eqn:ex33cd}
    \begin{align}
        \label{eqn:ex33c}
        \mhq &=   \rme^{-\rmi q_1}  
        \begin{pmatrix}
            -c_2 & c_1 + c_3 & 0 \\
            c_4 & 0 & c_1 +c_3 \\
            0 & c_4 & c_2 \\
        \end{pmatrix}
        + \mathrm{h.c.},\\
        \label{eqn:ex33d}
        \mhq &=   \rme^{-\rmi q_1}  
        \begin{pmatrix}
            0 & c_1 + c_3 & 0 \\
            -\frac{c_5 \bar{c}_3}{\bar{c}_4} & c_2 & c_1 +c_4 \\
            0 & c_5 & 0 \\
        \end{pmatrix}
        + \mathrm{h.c.}
    \end{align}
\end{subequations}
are flat-band Hamiltonians in 1D. Here, we took $q_1 = q_2 = q_3$. Taking $q_2 = 2q_1$ and $q_3 = 3 q_1$ would correspond to a one-dimensional lattice with three bands, nearest and next-nearest neighbour hopping (Fig.~\ref{fig:3}).

\subsection{$U > 1$}
\label{sec:UN}

As another example, we use the necessary and sufficient conditions~\eqref{eqn:LPFBgen-Nbands-syst-complex} to find new flat band classes with $U > 1$, for which a complete generator is currently missing. If $\EFB \neq b_n$ for all $n = 1, \ldots, N$ (the elements of $H_0$), then the corresponding flat band cannot have CLSs in the class $U = 1$. This follows from the condition~\eqref{eqn:LPFBgen-Nbands-syst-1-complex-2}: if $U = 1$ then $\EFB$ is an eigenvalue of $H_0$ because the eigenproblem $\left(\mhq - \EFB\IM\right) \psi$ reduces to $H_0\psi = \EFB \psi$.

Instead of starting from a fully undetermined Hamiltonian by attempting to solve the conditions~\eqref{eqn:LPFBgen-Nbands-syst-complex} for all the parameters of the Hamiltonian, we start from the well-known lattice model of the diamond chain,
\begin{equation}
    \label{eqn:UNex1-0}
    \mhq = \rme^{-\rmi q_1}
    \begin{pmatrix}
        0 & 1 & 0 \\
        0 & 1 & 0 \\
        0 & 1 & 0 \\
    \end{pmatrix}
    + \rme^{\rmi q_1}
    \begin{pmatrix}
        0 & 0 & 0 \\
        1 & 1 & 1 \\
        0 & 0 & 0 \\
    \end{pmatrix}
    +
    \begin{pmatrix}
        0 & 1 & 0 \\
        1 & 0 & 1 \\
        0 & 1 & 0 \\
    \end{pmatrix},
\end{equation}
which has the gapped spectrum $\lbrace 0, -2, 2\left[ 1 + \cos{(q_1)} \right] \rbrace$. The flat band at $0$ belongs to the $U = 1$ class, while the flat band at $-2$ belongs to the $U = 2$ class. In the canonical basis where $H_0$ is diagonal, the Hamiltonian reads $\mhq = \rme^{-\rmi q_1}  H_1^\dagger + \rme^{\rmi q_1} H_1 + H_0$, where
\begin{subequations}
    \label{eqn:UNex1}
    \begin{align}
        H_1 &= \begin{pmatrix}
            \frac{1}{2}-\frac{1}{\sqrt{2}} & \frac{1}{\sqrt{2}}-\frac{1}{2} & 0 \\
            -\frac{1}{2}-\frac{1}{\sqrt{2}} & \frac{1}{2}+\frac{1}{\sqrt{2}} & 0 \\
            0 & 0 & 0 \\
        \end{pmatrix},\\
        \label{eqn:UNex1-H0}
        H_0 &= \begin{pmatrix}
            -\sqrt{2} & 0 & 0 \\
            0 & \sqrt{2} & 0 \\
            0 & 0 & 0 \\
        \end{pmatrix}.
    \end{align}
\end{subequations}
Indeed, the flat band $\EFB = 0$ coincides with the diagonal elements of $H_0$ and $H_1$ takes the form of the $U=1$ solution~\cite{flach2014detangling}, while the $\EFB = -2$ solution does not. The Hamiltonian~\eqref{eqn:UNex1} satisfies the conditions~\eqref{eqn:LPFBgen-Nbands-syst-complex} if and only if $\lambda = 0$ or $\lambda = -2$. To see this, the conditions~\eqref{eqn:LPFBgen-Nbands-syst-complex} with $\lambda$ as the only unknown reduce to the set $\lambda(\lambda + 2) = 0$ and $\lambda(\lambda^2 - 4) = 0$ that needs to be satisfied simultaneously. Our strategy in what follows is to use the conditions~\eqref{eqn:LPFBgen-Nbands-syst-complex} to see whether the Hamiltonian~\eqref{eqn:UNex1} can be generalised. We will allow selected matrix elements of $H_1$ to become variables, and then use the necessary and sufficient conditions to check whether there exists flat bands.

This analysis readily reveals the easy observation that the $\EFB = -2$ flat band is immune to the zero on the third diagonal element of $H_1$ in the canonical basis while the $\lambda = 0$ flat band is not: if we replace the zero by $J \in \mathbb{C}$ then the eigenvalues read $\left\{-2,2 [\cos (q_1)+1], 2\, \mathrm{Re}(\rme^{-\rmi q_1} J) \right\}$. On the other hand, we can also have one (but not both) of the off-diagonal corners of $H_1$ non-zero, which is much more difficult to infer directly by diagonalisation. Consequently, in the basis of Eq.~\eqref{eqn:UNex1-0}, we have $\mathcal{H}_q = 
 \rme^{-\rmi q_1}  \tilde{H}_1^\dagger + \rme^{\rmi q_1} \tilde{H}_1 + \tilde{H}_0$, where
 \begin{subequations}
\label{eqn:UNex1-2}
\begin{align}
\tilde{H}_1^{\mathrm{(a)}} &=   \begin{pmatrix}
 -\frac{J}{4} &1+\frac{J}{2 \sqrt{2}}& -\frac{J}{4} \\
 0 & 1 & 0 \\
 \frac{J}{4} & 1-\frac{J}{2 \sqrt{2}} & \frac{J}{4} \\
\end{pmatrix},\\
\tilde{H}_1^{\mathrm{(b)}} &=   \begin{pmatrix}
 -\frac{J}{4} & 1 & \frac{J}{4} \\
 \frac{J}{2\sqrt{2}} & 1 & -\frac{J}{2\sqrt{2}} \\
 -\frac{J}{4} & 1 & \frac{J}{4} \\
\end{pmatrix},\\
\tilde{H}_0 &=  \begin{pmatrix}
0 & 1 & 0 \\
1 & 0 & 1 \\
0 & 1 & 0 \\
\end{pmatrix}.
\end{align}
\end{subequations}
Here there are two equivalent choices for $\tilde{H}_1$, labelled by `a' and `b'. The necessary and sufficient conditions~\eqref{eqn:LPFBgen-Nbands-syst-complex} show that there is a flat band at $\lambda = 2 (1 + \sqrt{2})$, distinct from all the diagonal elements of $H_0$ [Eq.~\eqref{eqn:UNex1-H0}], if and only if $J = 2 \sqrt{2 \left(\sqrt{2}+2\right)} \exp (\rmi \theta )$, where $\theta$ is real and arbitrary. It can be shown by e.g. direct numerical diagonalisation that $\theta$ does not have to be a constant across the system and, remarkably, the existence of the $U > 1$ flat band is immune to any texture or disorder in $\theta$.  If $J = 0$ we recover the lattice~\eqref{eqn:UNex1}. These cases exhaust the possibilites for flat bands in the system~\eqref{eqn:UNex1-2}.

\section{\label{sec:dc}Discussion and Conclusions}

\begin{figure}[t]
  \centering
        \begin{tikzpicture}
        \def\x{4.5};        \def\y{3.5};
              \node at (0,0) {    \includegraphics[width=0.23\textwidth,angle=0]{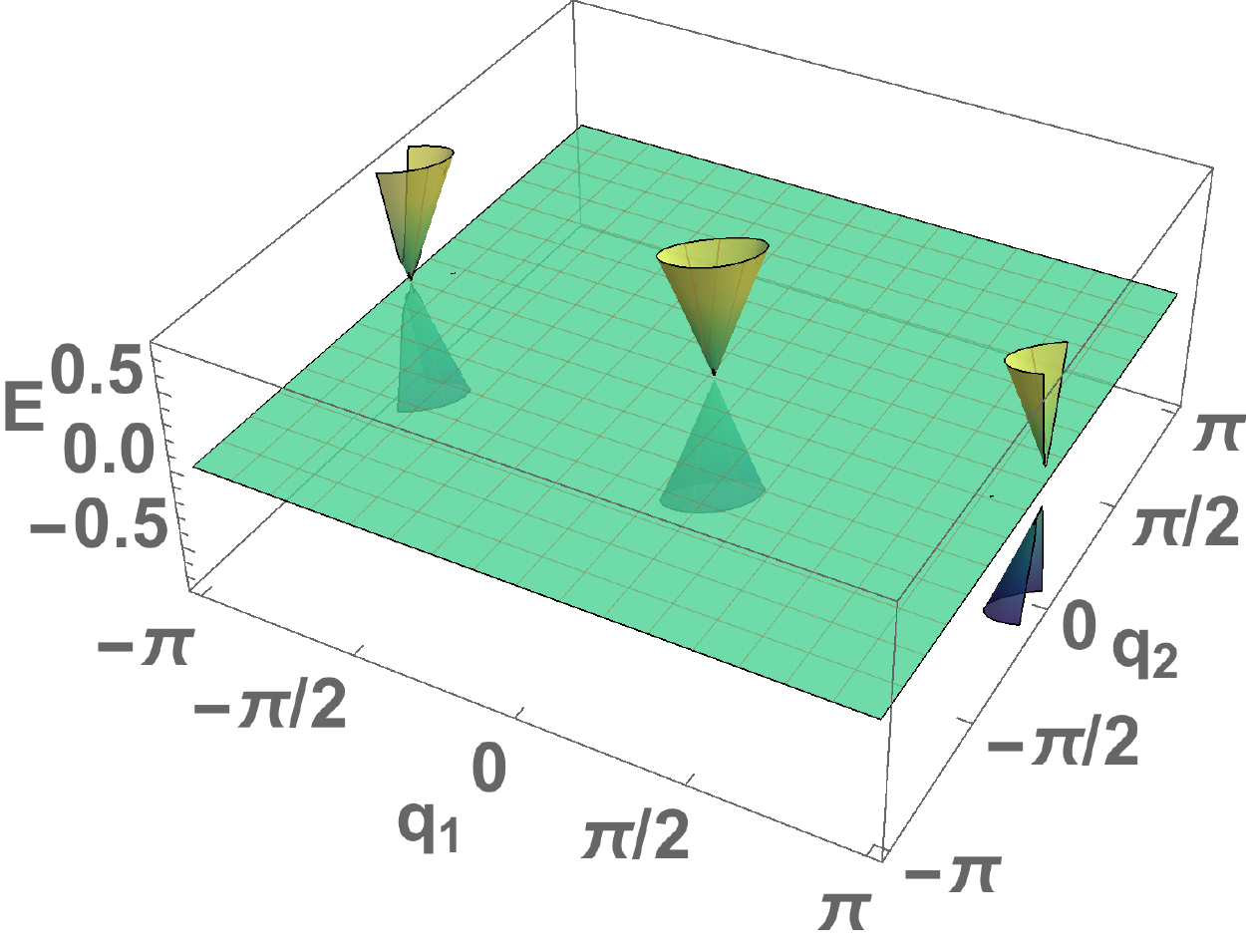}};
              \node at (\x,0) {    \includegraphics[width=0.23\textwidth,angle=0]{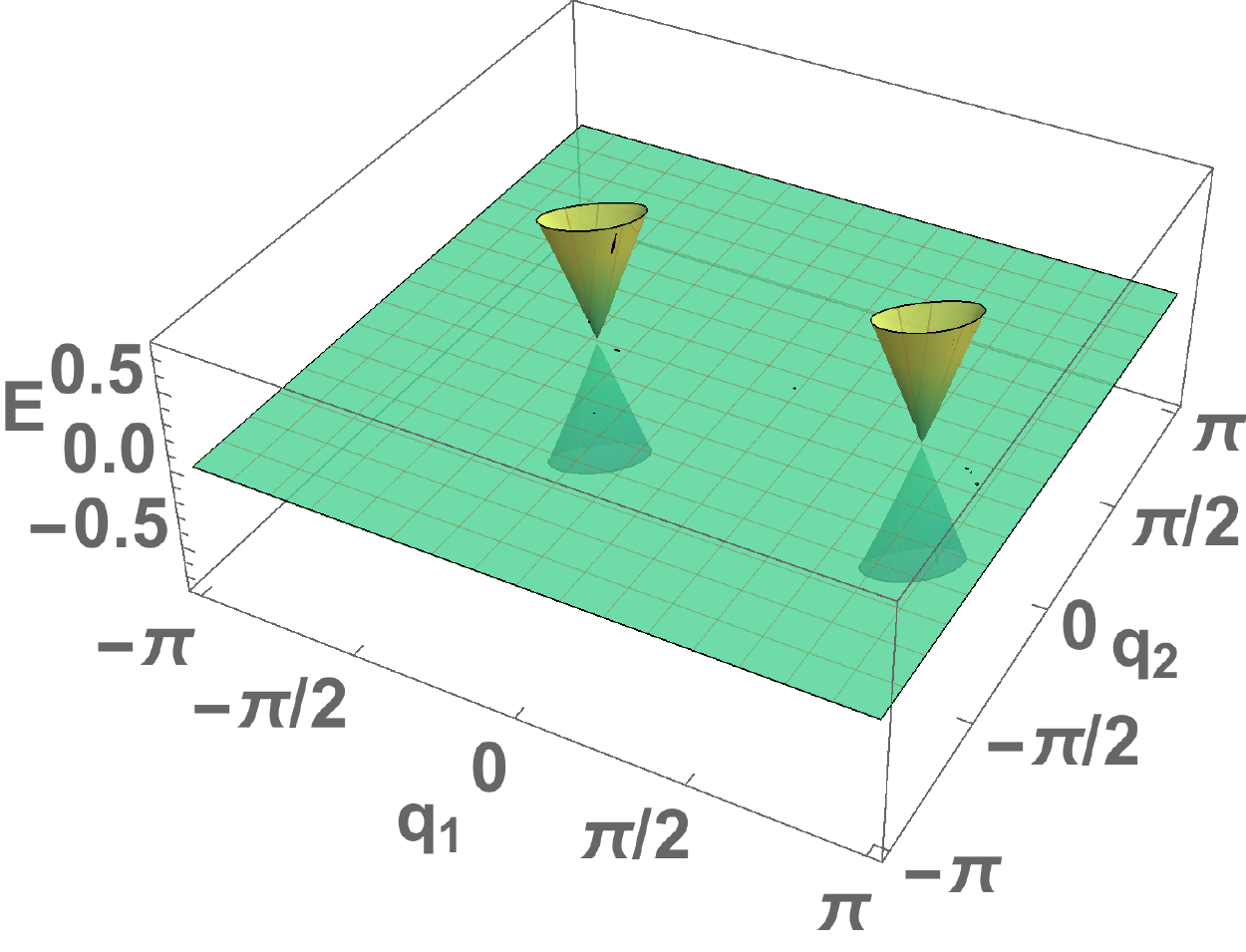}};
	    \node at (-1.5,1.3) {(a)};
            \node at (-1.5+\x,1.3) {(b)};
      \end{tikzpicture}    

      \caption{\label{fig:5}\textit{Weyl semi-metal with a flat band}. Eigenvalues of the Hamiltonian~\eqref{eqn:ex33a} with $q_3 = q_1 - q_2$, $c_1 = 1$, $c_2 = \rme^{\rmi \frac{\pi}{2}}$, $c_3 = \rme^{\rmi \frac{4\pi}{3}}$, and (a) $c_4 = \rme^{-\rmi \frac{\pi^2}{5} }$; (b) $c_4 = \rme^{\rmi \frac{\pi^2}{5} }$. The parameter $c_4$ controls the position of the Weyl cones along the $q_1$-direction.  }
\end{figure}

Recent attempts at classifying and generating flat bands have concentrated on identifying the parameter $U$, the minimum number of unit cells that a compact localized state associated with the flat band occupies, or lattice-specific methods that lack generality. Here, by focussing on the momentum-space representation, we have used an entirely different approach which allowed us to derive the full set of sufficient and necessary existence conditions for flat bands in generic tight-binding lattices. Our approach, using the necessary and sufficient conditions~\eqref{eqn:LPFBgen-Nbands-syst-complex-Mdim}, is fully distinct, and it does not specify the real-space property $U$ explicitly. Indeed, solving the conditions for parametrised Hamiltonians is a powerful method for generating new flat band lattices with arbitrary $U$, representing a novel and highly general flat band generator. Moreover, our new flat band existence conditions can serve as an important guide when generalising approaches that focus on labelling flat band states in terms of $U$, and finding lattices with $1 \leq n \leq N$ flat bands. Importantly, any flat band generator using any methods must be simultaneously consistent with our general existence conditions.

An immediate benefit of the explicit system~\eqref{eqn:LPFBgen-Nbands-syst-complex-Mdim} is that it allows us to directly and systematically generate non-trivial flat band Hamiltonians, especially for the $U > 1$ classes for which a fully general generator is currently unknown. Direct analytical diagonalisation of a `guessed' Hamiltonian with free parameters is not feasible for $N > 2$ and by the Abel-Ruffini theorem even impossible in general for $N > 4$. Our necessary and sufficient conditions, on the other hand, exhaust the possibilities for flat bands and can be straightforwardly computed for any $N > 4$. Of course, solving the necessary and sufficient existence conditions becomes challenging as a function of the complexity of the system, which may ultimately require a detailed numerical approach. We present results for $N = 4, 5$ in Appendix~\ref{app:C}, for $U > 1$ in Sec.~\ref{sec:UN} using the one-dimensional conditions~\eqref{eqn:LPFBgen-Nbands-syst-complex}, and for $U = 1$ in Sec.~\ref{sec:U1} using the more general conditions~\eqref{eqn:LPFBgen-Nbands-syst-complex-Mdim}. Moreover, we can consider flat-band systems with beyond-nearest-neighbour interactions between unit cells.

Indeed, the range of examples of new flat band systems that can be generated using our approach is substantial. One of the examples we found in Sec.~\ref{sec:U1} resembles an interesting class of materials called a Weyl semi-metal. The band structure of this exotic phase of matter is characterised by a pair of points in momentum space where two non-degenerate bands touch~\cite{wan2011topological}. The Weyl points are doubly degenerate, carry a quantized charge of Berry curvature resulting in a definite chirality, and are distinct from the four-fold degenerate chirality-neutral Dirac points associated with the band structure of e.g. graphene. The flat band Hamiltonian~\eqref{eqn:ex33a} with $q_3 = q_1 - q_2$ shows such a structure with suitable lattice parameters (Fig.~\ref{fig:5}). The location of the two Weyl cones can be adjusted by varying the free parameters of the Hamiltonian~\eqref{eqn:ex33a}. While not dissimilar from a Weyl semi-metal, our band structure is distinct in that it is supplemented by an additional flat band in the plane of the Weyl nodes.

In summary, we have derived the necessary and sufficient conditions for the existence of flat bands in $M$-dimensional tight-binding lattices with $N$ sites per unit cell and complex-amplitude nearest-neighbour tunneling between the lattice sites. That they are sufficient is trivial, and that they are also necessary can be proven by induction. If and only if the conditions are satisfied, then the system has one or more flat bands.

\acknowledgements
This work was supported by the Institute for Basic Science in Korea (IBS-R024-D1) and Austrian Academy of Sciences (P7050-029-011). 

\onecolumngrid
\appendix

\section{\label{app:B}The functions $\tilde{f}$ explicitly for $N \leq 4$ with $H_1 \in \mathbb{C}$}
 \begin{subequations}
\label{eqn:LPFBgen-Nbands-syst-addprop1-complex}
\begin{align}
\label{eqn:LPFBgen-Nbands-syst-addprop1-21-complex}
\tilde{f}_{-1}^{(2)}(\lambda,H_0,H_1) &= \det{\left(H_0 - \lambda I_2 \right)} -  \mathrm{tr}{\left(H_1 C_0 \right)} =  \sum_{ ij \in \mathcal{P}_2} \left(\frac{ \lambda^2}{2} - \left(\lambda + h_{jj}\right) (H_0)_{ii} +  \frac{(H_0)_{ii}(H_0)_{jj}}{2}  \right),\\
\label{eqn:LPFBgen-Nbands-syst-addprop1-31-complex} 
\tilde{f}_{-1}^{(3)}(\lambda,H_0,H_1)
 &=  \det{\left(H_0 - \lambda I_3 \right)} + \sum_{ ijk \in \mathcal{P}_3}  \frac{(H_0 - \lambda I_3)_{ii}}{2} g(j,k),     \\
\label{eqn:LPFBgen-Nbands-syst-addprop1-41-complex} 
\tilde{f}_{-1}^{(4)}(\lambda,H_0,H_1) &= \det{\left(H_0 - \lambda I_4 \right)} + \sum_{ ijkl \in \mathcal{P}_4} \left[ \frac{\lambda^2}{4}  + \frac{(H_0)_{ll}}{4} (H_0 - \lambda I_4)_{ii}  \right]g(j,k) + \tilde{f}_{\mathrm{1a}}^{(4)}(H_1) + \tilde{f}_{\mathrm{1b}}^{(4)}(H_1),  \\
%
\notag \tilde{f}_{\mathrm{1a}}^{(4)}(H_1) &= \sum_{ ijkl \in \mathcal{P}_4}\left\lbrace \frac{1}{8}g(i,l) g(j,k) + \mathrm{Re}\left[ - h_{ij}^*h_{jk}^*h_{il}h_{lk} 
+2h_{ij}^*h_{jk}^*h_{ik}h_{ll} -\frac{1}{2}h_{ij}^*h_{ik}^*h_{lj}h_{lk} \right. \right. \\
\notag & \left. \left. \qquad \qquad\qquad
+\frac{1}{4}h_{ij}^*h_{ji}^*h_{kl}h_{lk}
-\frac{1}{2}h_{ij}^*h_{ji}^*h_{kk}h_{ll}
-\frac{1}{4}h_{ii}^*h_{jj}^*h_{kk}h_{ll} \right]
 \right\rbrace,\\
 \notag \tilde{f}_{\mathrm{1b}}^{(4)}(H_1) 
 &=\rmi \sum_{ ijkl \in \mathcal{P}_4}\mathrm{Im} \left(
h_{kj}h_{ij}^*
 \right)h_{kl} h_{il}^*  , \\
\label{eqn:LPFBgen-Nbands-syst-addprop1-32-complex}
\tilde{f}_{-2}^{(3)}(\lambda,H_0,H_1) &=  - \frac{\lambda}{2}   \sum_{ ijk \in \mathcal{P}_3}\left(2 H_0 - \lambda I_3 \right)_{jj}  h_{ii} , \\
\label{eqn:LPFBgen-Nbands-syst-addprop1-42-complex}
\tilde{f}_{-2}^{(4)}(\lambda,H_0,H_1) &= \sum_{ ijkl \in \mathcal{P}_4} \left\lbrace
-\frac{\lambda^3}{6} h_{ii}^*
+ \frac{\lambda^2}{2} (H_0)_{ii}h_{jj}^* +\frac{1}{6} (H_0)_{ii}(H_0)_{jj} (H_0)_{kk} h_{ll}^*
\right. \\
\notag & +\lambda \left[ 
- \frac{1}{2} (H_0)_{ii}(H_0)_{jj}h_{kk}^*
-\frac{1}{2} h_{ii}h_{jj}^*h_{kk}^*
+ h_{ij}h_{ij}^* h_{kk}^*
-h_{ij}h_{ik}^* h_{kj}^*
+ \frac{1}{2} h_{ii} h_{jk}^* h_{kj}^*
 \right] \\
 \notag & \left. 
 + (H_0)_{ii} \left[
 \frac{1}{2}h_{ll} h_{jj}^*  h_{kk}^* 
 -h_{lk}  h_{kk}^*  h_{lj}^*  
 -\frac{1}{2} h_{ll}  h_{jk}^*  h_{kj}^*  
 + h_{kl}  h_{jl}^*  h_{kj}^*    
 \right]
 \right\rbrace ,\\
\label{eqn:LPFBgen-Nbands-syst-addprop1-43-complex}
\tilde{f}_{-3}^{(4)}(\lambda,H_0,H_1) &=\sum_{ ijkl \in \mathcal{P}_4}\left\lbrace \left(\frac{\lambda^2}{4} - \frac{\lambda}{2}  (H_0)_{ii} + \frac{(H_0)_{ii}(H_0)_{ll}}{4} \right) \left(- h_{jk}h_{kj} + h_{jj}h_{kk} \right)    
-\frac{1}{6}(H_0)_{ii}(H_0)_{jj}(H_0)_{kk}h_{ll}
\right\rbrace ,\\
\notag g(j,k) &=  -\left|\left(H_1\right)_{jk}\right|^2  -\left|\left(H_1\right)_{kj}\right|^2 + \left(H_1\right)_{kk} \left(H_1\right)_{jj}^* +\left(H_1\right)_{kk}^* \left(H_1\right)_{jj}    ,\\
\notag h_{ij} &=  \left(H_1  \right)_{ij},
\end{align}
\end{subequations}
where $ijk \in \mathcal{P}_3$ means all the six permutations of $(1,2,3)$.

\section{\label{app:A}The only non-zero cross and $q$-independent terms for specific $(N,M)$ }
We introduce the notation $H_1^{(m)} \equiv H_{1,m}$.
\begin{itemize}
\item $N = 2$ and $M = 2$
\begin{subequations}
\begin{align}
\bar{y}_{1,1} ^{(2,2)} = y_{-1,-1} ^{(2,2)} &= h_{11}^{(1)}h_{22}^{(2)}  - h_{12}^{(1)} h_{21}^{(2)} + h_{22}^{(1)} h_{11}^{(2)} - h_{21}^{(1)} h_{12}^{(2)} , \\
\bar{y}_{-1,1} ^{(2,2)} = y_{1,-1} ^{(2,2)} &= h^{(2)}_{11} \bar{h}^{(1)}_{22}  + h^{(2)}_{22}\bar{h}^{(1)}_{11} -h^{(2)}_{12} \bar{h}^{(1)}_{12}-h^{(2)}_{21} \bar{h}^{(1)}_{21},\\
F_0^{(2,2)} &= \det{\left(H_0 - \lambda I_2 \right)} 
+h^{(1)}_{22} \bar{h}^{(1)}_{11}+h^{(1)}_{11} \bar{h}^{(1)}_{22}-h^{(1)}_{12} \bar{h}^{(1)}_{12}\\
\notag & \;\;\;-h^{(1)}_{21} \bar{h}^{(1)}_{21}+h^{(2)}_{22} \bar{h}^{(2)}_{11}+h^{(2)}_{11} \bar{h}^{(2)}_{22}-h^{(2)}_{12} \bar{h}^{(2)}_{12}-h^{(2)}_{21} \bar{h}^{(2)}_{21}.
\end{align}
\end{subequations}

\item $N = 2$ and $M = 3$
\begin{subequations}
\begin{align}
\bar{y}_{0,1,1} ^{(2,3)} = y_{0,-1,-1} ^{(2,3)} &= h^{(2)}_{11} h^{(3)}_{22}-h^{(2)}_{12} h^{(3)}_{21}-h^{(2)}_{21} h^{(3)}_{12}+h^{(2)}_{22} h^{(3)}_{11}, \\
\bar{y}_{1,0,1} ^{(2,3)} = y_{-1,0,-1} ^{(2,3)} &= h^{(1)}_{11} h^{(3)}_{22}-h^{(1)}_{12} h^{(3)}_{21}-h^{(1)}_{21} h^{(3)}_{12}+h^{(1)}_{22} h^{(3)}_{11}, \\
\bar{y}_{1,1,0} ^{(2,3)} = y_{-1,-1,0} ^{(2,3)} &= h^{(1)}_{11} h^{(2)}_{22}-h^{(1)}_{12} h^{(2)}_{21}-h^{(1)}_{21} h^{(2)}_{12}+h^{(1)}_{22} h^{(2)}_{11}, \\
\bar{y}_{-1,0,1} ^{(2,3)} = y_{1,0,-1} ^{(2,3)} &= h^{(3)}_{22} \bar{h}^{(1)}_{11}-h^{(3)}_{12} \bar{h}^{(1)}_{12}-h^{(3)}_{21} \bar{h}^{(1)}_{21}+h^{(3)}_{11} \bar{h}^{(1)}_{22},\\
\bar{y}_{-1,1,0} ^{(2,3)} = y_{1,-1,0} ^{(2,3)} &= h^{(2)}_{22} \bar{h}^{(1)}_{11} -h^{(2)}_{12} \bar{h}^{(1)}_{12} -h^{(2)}_{21} \bar{h}^{(1)}_{21}+h^{(2)}_{11} \bar{h}^{(1)}_{22},\\
\bar{y}_{0,-1,1} ^{(2,3)} = y_{0,1,-1} ^{(2,3)} &=h^{(3)}_{22} \bar{h}^{(2)}_{11} -h^{(3)}_{12} \bar{h}^{(2)}_{12} -h^{(3)}_{21} \bar{h}^{(2)}_{21} +h^{(3)}_{11} \bar{h}^{(2)}_{22},\\
F_0^{(2,3)} &= F_0^{(2,2)} + h^{(3)}_{22} \bar{h}^{(3)}_{11}+h^{(3)}_{11} \bar{h}^{(3)}_{22}-h^{(3)}_{12} \bar{h}^{(3)}_{12}-h^{(3)}_{21} \bar{h}^{(3)}_{21}.
\end{align}
\end{subequations}

\item $N = 3$ and $M = 2$
\begin{subequations}
\begin{align}
\bar{y}_{2,1} ^{(3,2)} = y_{-2,-1} ^{(3,2)} &= h^{(1)}_{11} h^{(1)}_{22} h^{(2)}_{33}-h^{(1)}_{11} h^{(1)}_{23} h^{(2)}_{32}-h^{(1)}_{11} h^{(1)}_{32} h^{(2)}_{23}+h^{(1)}_{11} h^{(1)}_{33} h^{(2)}_{22}  -h^{(1)}_{12} h^{(1)}_{21} h^{(2)}_{33}+h^{(1)}_{12} h^{(1)}_{23} h^{(2)}_{31}\\
\notag &+h^{(1)}_{12} h^{(1)}_{31} h^{(2)}_{23}-h^{(1)}_{12} h^{(1)}_{33} h^{(2)}_{21}  +h^{(1)}_{13} h^{(1)}_{21} h^{(2)}_{32}-h^{(1)}_{13} h^{(1)}_{22} h^{(2)}_{31}-h^{(1)}_{13} h^{(1)}_{31} h^{(2)}_{22}+h^{(1)}_{13} h^{(1)}_{32} h^{(2)}_{21} \\
\notag & +h^{(1)}_{21} h^{(1)}_{32} h^{(2)}_{13}-h^{(1)}_{21} h^{(1)}_{33} h^{(2)}_{12}-h^{(1)}_{22} h^{(1)}_{31} h^{(2)}_{13}+h^{(1)}_{22} h^{(1)}_{33} h^{(2)}_{11} +h^{(1)}_{23} h^{(1)}_{31} h^{(2)}_{12}-h^{(1)}_{23} h^{(1)}_{32} h^{(2)}_{11}, \\
\bar{y}_{2,-1} ^{(3,2)} = y_{-2,1} ^{(3,2)} &=
h^{(1)}_{11} h^{(1)}_{22} \bar{h}^{(2)}_{33}-h^{(1)}_{11} h^{(1)}_{23} \bar{h}^{(2)}_{23}-h^{(1)}_{11} h^{(1)}_{32} \bar{h}^{(2)}_{32}+h^{(1)}_{11} h^{(1)}_{33} \bar{h}^{(2)}_{22} -h^{(1)}_{12} h^{(1)}_{21} \bar{h}^{(2)}_{33}+h^{(1)}_{12} h^{(1)}_{23} \bar{h}^{(2)}_{13}\\
\notag &+h^{(1)}_{12} h^{(1)}_{31} \bar{h}^{(2)}_{32}-h^{(1)}_{12} h^{(1)}_{33} \bar{h}^{(2)}_{12} +h^{(1)}_{13} h^{(1)}_{21} \bar{h}^{(2)}_{23}-h^{(1)}_{13} h^{(1)}_{22} \bar{h}^{(2)}_{13}-h^{(1)}_{13} h^{(1)}_{31} \bar{h}^{(2)}_{22}+h^{(1)}_{13} h^{(1)}_{32} \bar{h}^{(2)}_{12}\\
\notag &+h^{(1)}_{21} h^{(1)}_{32} \bar{h}^{(2)}_{31}-h^{(1)}_{21} h^{(1)}_{33} \bar{h}^{(2)}_{21}-h^{(1)}_{22} h^{(1)}_{31} \bar{h}^{(2)}_{31}+h^{(1)}_{22} h^{(1)}_{33} \bar{h}^{(2)}_{11} +h^{(1)}_{23} h^{(1)}_{31} \bar{h}^{(2)}_{21}-h^{(1)}_{23} h^{(1)}_{32} \bar{h}^{(2)}_{11},\\
\bar{y}_{1,2} ^{(3,2)} = y_{-1,-2} ^{(3,2)} &=
h^{(1)}_{11} h^{(2)}_{22} h^{(2)}_{33}-h^{(1)}_{11} h^{(2)}_{23} h^{(2)}_{32}-h^{(1)}_{12} h^{(2)}_{21} h^{(2)}_{33}+h^{(1)}_{12} h^{(2)}_{23} h^{(2)}_{31}+h^{(1)}_{13} h^{(2)}_{21} h^{(2)}_{32}-h^{(1)}_{13} h^{(2)}_{22} h^{(2)}_{31}\\
\notag &-h^{(1)}_{21} h^{(2)}_{12} h^{(2)}_{33}+h^{(1)}_{21} h^{(2)}_{13} h^{(2)}_{32} +h^{(1)}_{22} h^{(2)}_{11} h^{(2)}_{33}-h^{(1)}_{22} h^{(2)}_{13} h^{(2)}_{31}-h^{(1)}_{23} h^{(2)}_{11} h^{(2)}_{32}+h^{(1)}_{23} h^{(2)}_{12} h^{(2)}_{31}\\
\notag &+h^{(1)}_{31} h^{(2)}_{12} h^{(2)}_{23}-h^{(1)}_{31} h^{(2)}_{13} h^{(2)}_{22}-h^{(1)}_{32} h^{(2)}_{11} h^{(2)}_{23}+h^{(1)}_{32} h^{(2)}_{13} h^{(2)}_{21}+h^{(1)}_{33} h^{(2)}_{11} h^{(2)}_{22}-h^{(1)}_{33} h^{(2)}_{12} h^{(2)}_{21},\\
\bar{y}_{1,1} ^{(3,2)} = y_{-1,-1} ^{(3,2)} &=
h^{(1)}_{11} h^{(2)}_{22} (H_0)_{33}-h^{(1)}_{11} h^{(2)}_{22} \lambda +h^{(1)}_{11} h^{(2)}_{33} (H_0)_{22}-h^{(1)}_{11} h^{(2)}_{33} \lambda -h^{(1)}_{12} h^{(2)}_{21} (H_0)_{33}+h^{(1)}_{12} h^{(2)}_{21} \lambda \\
\notag &-h^{(1)}_{13} h^{(2)}_{31} (H_0)_{22}+h^{(1)}_{13} h^{(2)}_{31} \lambda -h^{(1)}_{21} h^{(2)}_{12} (H_0)_{33}+h^{(1)}_{21} h^{(2)}_{12} \lambda +h^{(1)}_{22} h^{(2)}_{11} (H_0)_{33}-h^{(1)}_{22} h^{(2)}_{11} \lambda \\
\notag & +h^{(1)}_{22} h^{(2)}_{33} (H_0)_{11}-h^{(1)}_{22} h^{(2)}_{33} \lambda -h^{(1)}_{23} h^{(2)}_{32} (H_0)_{11}+h^{(1)}_{23} h^{(2)}_{32} \lambda -h^{(1)}_{31} h^{(2)}_{13} (H_0)_{22}+h^{(1)}_{31} h^{(2)}_{13} \lambda\\
\notag &  -h^{(1)}_{32} h^{(2)}_{23} (H_0)_{11}+h^{(1)}_{32} h^{(2)}_{23} \lambda  +h^{(1)}_{33} h^{(2)}_{11} (H_0)_{22}-h^{(1)}_{33} h^{(2)}_{11} \lambda +h^{(1)}_{33} h^{(2)}_{22} (H_0)_{11}-h^{(1)}_{33} h^{(2)}_{22} \lambda,\\
\bar{y}_{1,-1} ^{(3,2)} = y_{-1,1} ^{(3,2)} &=
h^{(1)}_{11} (H_0)_{33} \bar{h}^{(2)}_{22}-h^{(1)}_{11} \lambda  \bar{h}^{(2)}_{22}+h^{(1)}_{11} (H_0)_{22} \bar{h}^{(2)}_{33}-h^{(1)}_{11} \lambda  \bar{h}^{(2)}_{33}-h^{(1)}_{12} (H_0)_{33} \bar{h}^{(2)}_{12}+h^{(1)}_{12} \lambda  \bar{h}^{(2)}_{12}\\
\notag &-h^{(1)}_{13} (H_0)_{22} \bar{h}^{(2)}_{13}+h^{(1)}_{13} \lambda  \bar{h}^{(2)}_{13}-h^{(1)}_{21} (H_0)_{33} \bar{h}^{(2)}_{21}+h^{(1)}_{21} \lambda  \bar{h}^{(2)}_{21}+h^{(1)}_{22} (H_0)_{33} \bar{h}^{(2)}_{11}-h^{(1)}_{22} \lambda  \bar{h}^{(2)}_{11}\\
\notag &+h^{(1)}_{22} (H_0)_{11} \bar{h}^{(2)}_{33}-h^{(1)}_{22} \lambda  \bar{h}^{(2)}_{33}-h^{(1)}_{23} (H_0)_{11} \bar{h}^{(2)}_{23}+h^{(1)}_{23} \lambda  \bar{h}^{(2)}_{23}-h^{(1)}_{31} (H_0)_{22} \bar{h}^{(2)}_{31}+h^{(1)}_{31} \lambda  \bar{h}^{(2)}_{31}\\
\notag &-h^{(1)}_{32} (H_0)_{11} \bar{h}^{(2)}_{32}+h^{(1)}_{32} \lambda  \bar{h}^{(2)}_{32} +h^{(1)}_{33} (H_0)_{22} \bar{h}^{(2)}_{11}-h^{(1)}_{33} \lambda  \bar{h}^{(2)}_{11}+h^{(1)}_{33} (H_0)_{11} \bar{h}^{(2)}_{22}-h^{(1)}_{33} \lambda  \bar{h}^{(2)}_{22},\\
\bar{y}_{-1,2} ^{(3,2)} = y_{1,-2} ^{(3,2)} &=
h^{(2)}_{22} h^{(2)}_{33} \bar{h}^{(1)}_{11}-h^{(2)}_{23} h^{(2)}_{32} \bar{h}^{(1)}_{11}-h^{(2)}_{12} h^{(2)}_{33} \bar{h}^{(1)}_{12}+h^{(2)}_{13} h^{(2)}_{32} \bar{h}^{(1)}_{12}+h^{(2)}_{12} h^{(2)}_{23} \bar{h}^{(1)}_{13}-h^{(2)}_{13} h^{(2)}_{22} \bar{h}^{(1)}_{13}\\
\notag &-h^{(2)}_{21} h^{(2)}_{33} \bar{h}^{(1)}_{21}+h^{(2)}_{23} h^{(2)}_{31} \bar{h}^{(1)}_{21} +h^{(2)}_{11} h^{(2)}_{33} \bar{h}^{(1)}_{22}-h^{(2)}_{13} h^{(2)}_{31} \bar{h}^{(1)}_{22}-h^{(2)}_{11} h^{(2)}_{23} \bar{h}^{(1)}_{23}+h^{(2)}_{13} h^{(2)}_{21} \bar{h}^{(1)}_{23}\\
\notag &+h^{(2)}_{21} h^{(2)}_{32} \bar{h}^{(1)}_{31}-h^{(2)}_{22} h^{(2)}_{31} \bar{h}^{(1)}_{31}-h^{(2)}_{11} h^{(2)}_{32} \bar{h}^{(1)}_{32}+h^{(2)}_{12} h^{(2)}_{31} \bar{h}^{(1)}_{32}+h^{(2)}_{11} h^{(2)}_{22} \bar{h}^{(1)}_{33}-h^{(2)}_{12} h^{(2)}_{21} \bar{h}^{(1)}_{33},\\
\bar{\hat{x}}^{(1,3)}_{1}=\hat{x}^{(1,3)}_{-1} &= 
h^{(1)}_{11} h^{(2)}_{33} \bar{h}^{(2)}_{22}+h^{(1)}_{11} h^{(2)}_{22} \bar{h}^{(2)}_{33}-h^{(1)}_{11} h^{(2)}_{23} \bar{h}^{(2)}_{23}-h^{(1)}_{11} h^{(2)}_{32} \bar{h}^{(2)}_{32}-h^{(1)}_{12} h^{(2)}_{33} \bar{h}^{(2)}_{12}+h^{(1)}_{12} h^{(2)}_{23} \bar{h}^{(2)}_{13}\\
\notag &
-h^{(1)}_{12} h^{(2)}_{21} \bar{h}^{(2)}_{33}+h^{(1)}_{12} h^{(2)}_{31}  \bar{h}^{(2)}_{32}+h^{(1)}_{13} h^{(2)}_{32}  \bar{h}^{(2)}_{12}-h^{(1)}_{13} h^{(2)}_{22}  \bar{h}^{(2)}_{13}+h^{(1)}_{13} h^{(2)}_{21}  \bar{h}^{(2)}_{23}-h^{(1)}_{13} h^{(2)}_{31}  \bar{h}^{(2)}_{22}\\
\notag &
-h^{(1)}_{21} h^{(2)}_{12} \bar{h}^{(2)}_{33} +h^{(1)}_{21} h^{(2)}_{13} \bar{h}^{(2)}_{23} -h^{(1)}_{21} h^{(2)}_{33} \bar{h}^{(2)}_{21} +h^{(1)}_{21} h^{(2)}_{32} \bar{h}^{(2)}_{31}+h^{(1)}_{22} h^{(2)}_{33} \bar{h}^{(2)}_{11}+h^{(1)}_{22} h^{(2)}_{11} \bar{h}^{(2)}_{33}\\
\notag &
-h^{(1)}_{22} h^{(2)}_{13} \bar{h}^{(2)}_{13} -h^{(1)}_{22} h^{(2)}_{31} \bar{h}^{(2)}_{31} -h^{(1)}_{23} h^{(2)}_{11} \bar{h}^{(2)}_{23} -h^{(1)}_{23} h^{(2)}_{32} \bar{h}^{(2)}_{11} +h^{(1)}_{23} h^{(2)}_{12} \bar{h}^{(2)}_{13} +h^{(1)}_{23} h^{(2)}_{31} \bar{h}^{(2)}_{21} \\
\notag &
+h^{(1)}_{31} h^{(2)}_{12} \bar{h}^{(2)}_{32} -h^{(1)}_{31} h^{(2)}_{13} \bar{h}^{(2)}_{22} +h^{(1)}_{31} h^{(2)}_{23} \bar{h}^{(2)}_{21} -h^{(1)}_{31} h^{(2)}_{22} \bar{h}^{(2)}_{31}-h^{(1)}_{32} h^{(2)}_{23} \bar{h}^{(2)}_{11} -h^{(1)}_{32} h^{(2)}_{11} \bar{h}^{(2)}_{32}\\
\notag &
+h^{(1)}_{32} h^{(2)}_{13} \bar{h}^{(2)}_{12} +h^{(1)}_{32} h^{(2)}_{21} \bar{h}^{(2)}_{31} +h^{(1)}_{33} h^{(2)}_{22} \bar{h}^{(2)}_{11} +h^{(1)}_{33} h^{(2)}_{11} \bar{h}^{(2)}_{22} -h^{(1)}_{33} h^{(2)}_{12} \bar{h}^{(2)}_{12} -h^{(1)}_{33} h^{(2)}_{21} \bar{h}^{(2)}_{21} ,\\
\bar{\hat{x}}^{(2,3)}_{1}  = \hat{x}^{(2,3)}_{-1} &= h^{(1)}_{22} h^{(2)}_{33} \bar{h}^{(1)}_{11}+h^{(1)}_{11} h^{(2)}_{33} \bar{h}^{(1)}_{22}-h^{(1)}_{11} h^{(2)}_{23} \bar{h}^{(1)}_{23}-h^{(1)}_{23} h^{(2)}_{32} \bar{h}^{(1)}_{11}-h^{(1)}_{32} h^{(2)}_{23} \bar{h}^{(1)}_{11}-h^{(1)}_{11} h^{(2)}_{32} \bar{h}^{(1)}_{32}\\
\notag &
+h^{(1)}_{33} h^{(2)}_{22} \bar{h}^{(1)}_{11}+h^{(1)}_{11} h^{(2)}_{22} \bar{h}^{(1)}_{33}+h^{(1)}_{12} h^{(2)}_{23} \bar{h}^{(1)}_{13}+h^{(1)}_{13} h^{(2)}_{32} \bar{h}^{(1)}_{12}+h^{(1)}_{32} h^{(2)}_{13} \bar{h}^{(1)}_{12}+h^{(1)}_{12} h^{(2)}_{31} \bar{h}^{(1)}_{32}\\
\notag &
-h^{(1)}_{33} h^{(2)}_{12} \bar{h}^{(1)}_{12}-h^{(1)}_{12} h^{(2)}_{21} \bar{h}^{(1)}_{33}-h^{(1)}_{12} h^{(2)}_{33} \bar{h}^{(1)}_{12}-h^{(1)}_{22} h^{(2)}_{13} \bar{h}^{(1)}_{13}-h^{(1)}_{13} h^{(2)}_{31} \bar{h}^{(1)}_{22}+h^{(1)}_{23} h^{(2)}_{12} \bar{h}^{(1)}_{13}\\
\notag &
+h^{(1)}_{13} h^{(2)}_{21} \bar{h}^{(1)}_{23}-h^{(1)}_{13} h^{(2)}_{22} \bar{h}^{(1)}_{13}+h^{(1)}_{21} h^{(2)}_{13} \bar{h}^{(1)}_{23}+h^{(1)}_{23} h^{(2)}_{31} \bar{h}^{(1)}_{21}+h^{(1)}_{31} h^{(2)}_{23} \bar{h}^{(1)}_{21}+h^{(1)}_{21} h^{(2)}_{32} \bar{h}^{(1)}_{31}\\
\notag &
-h^{(1)}_{21} h^{(2)}_{12} \bar{h}^{(1)}_{33}-h^{(1)}_{33} h^{(2)}_{21} \bar{h}^{(1)}_{21}-h^{(1)}_{21} h^{(2)}_{33} \bar{h}^{(1)}_{21}-h^{(1)}_{31} h^{(2)}_{13} \bar{h}^{(1)}_{22}-h^{(1)}_{22} h^{(2)}_{31} \bar{h}^{(1)}_{31}+h^{(1)}_{33} h^{(2)}_{11} \bar{h}^{(1)}_{22}\\
\notag &
+h^{(1)}_{22} h^{(2)}_{11} \bar{h}^{(1)}_{33}-h^{(1)}_{23} h^{(2)}_{11} \bar{h}^{(1)}_{23}+h^{(1)}_{31} h^{(2)}_{12} \bar{h}^{(1)}_{32}+h^{(1)}_{32} h^{(2)}_{21} \bar{h}^{(1)}_{31}-h^{(1)}_{31} h^{(2)}_{22} \bar{h}^{(1)}_{31}-h^{(1)}_{32} h^{(2)}_{11} \bar{h}^{(1)}_{32},\\
F_0^{(3,2)} &= \sum_{m=1}^2 \tilde{f}_{-1}^{(3)}(\lambda,H_0,H_1^{(m)}) -\det{\left(H_0 - \lambda I_3 \right)}.
\end{align}
\end{subequations}

\item $N = 3$ and $M = 3$. Instead of giving the terms in explicit detail as above, we simply list the non-zero ones:
\begin{subequations}
\begin{align}
\text{Non-zero $y$}&: 
\left( \bar{y}_{2,1,0} ^{(3,3)} = y_{-2,-1,0} ^{(3,3)}  \right), 
\left( \bar{y}_{2,0,1} ^{(3,3)} = y_{-2,0,-1} ^{(3,3)}  \right), 
\left( \bar{y}_{1,2,0} ^{(3,3)} = y_{-1,-2,0} ^{(3,3)}  \right), 
\left( \bar{y}_{1,1,1} ^{(3,3)} = y_{-1,-1,-1} ^{(3,3)}  \right), 
\left( \bar{y}_{1,1,0} ^{(3,3)} = y_{-1,-1,0} ^{(3,3)}  \right), \\
\notag &
\left( \bar{y}_{1,1,-1} ^{(3,3)} = y_{-1,-1,1} ^{(3,3)}  \right), 
\left( \bar{y}_{1,0,2} ^{(3,3)}  = y_{-1,0,-2} ^{(3,3)} \right),
\left( \bar{y}_{1,0,1} ^{(3,3)} = y_{-1,0,-1} ^{(3,3)}  \right),  
\left( \bar{y}_{1,0,-1} ^{(3,3)} = y_{-1,0,1} ^{(3,3)}  \right), 
\left( \bar{y}_{1,0,-2} ^{(3,3)} = y_{-1,0,2} ^{(3,3)}  \right),\\
\notag &
\left( \bar{y}_{1,-1,1} ^{(3,3)} = y_{-1,1,-1} ^{(3,3)}  \right),
\left( \bar{y}_{1,-1,0} ^{(3,3)} = y_{-1,1,0} ^{(3,3)}  \right),
\left( \bar{y}_{1,-1,-1} ^{(3,3)} = y_{-1,1,1} ^{(3,3)}  \right),
\left( \bar{y}_{1,-2,0} ^{(3,3)} = y_{-1,2,0} ^{(3,3)}  \right),
\left( \bar{y}_{0,2,1} ^{(3,3)} = y_{0,-2,-1} ^{(3,3)}  \right),\\
\notag &
\left( \bar{y}_{0,2,-1} ^{(3,3)} = y_{0,-2,1} ^{(3,3)}  \right),
\left( \bar{y}_{0,1,2} ^{(3,3)} = y_{0,-1,-2} ^{(3,3)}  \right),
\left( \bar{y}_{0,1,1} ^{(3,3)} = y_{0,-1,-1} ^{(3,3)}  \right),
\left( \bar{y}_{0,1,-1} ^{(3,3)} = y_{0,-1,1} ^{(3,3)}  \right),
\left( \bar{y}_{0,1,-2} ^{(3,3)} = y_{0,-1,2} ^{(3,3)}  \right),\\
\notag &
\left( \bar{y}_{-2,1,0} ^{(3,3)} = y_{2,-1,0} ^{(3,3)}  \right), 
\left( \bar{y}_{-2,0,1} ^{(3,3)} = y_{2,0,-1} ^{(3,3)}  \right),\\
\text{Non-zero $\hat{x}$}&:  \left( \bar{\hat{x}}^{(1,3)}_{1}  = \hat{x}^{(1,3)}_{-1}  \right), \left( \bar{\hat{x}}^{(2,3)}_{1}  = \hat{x}^{(2,3)}_{-1}  \right), \left( \bar{\hat{x}}^{(3,3)}_{1}  = \hat{x}^{(3,3)}_{-1}  \right), \\
F_0^{(3,3)} &= \sum_{m=1}^3 \tilde{f}_{-1}^{(3)}(\lambda,H_0,H_1^{(m)}) -2\det{\left(H_0 - \lambda I_3 \right)}.
\end{align}
\end{subequations}

\end{itemize}

\section{\label{app:C}More $U = 1$ flat band cases from solving the necessary and sufficient conditions}
\begin{figure}[t]
  \centering
        \begin{tikzpicture}
        \def\x{4.5};        \def\y{3.5};
              \node at (0,0) {    \includegraphics[width=0.23\textwidth,angle=0]{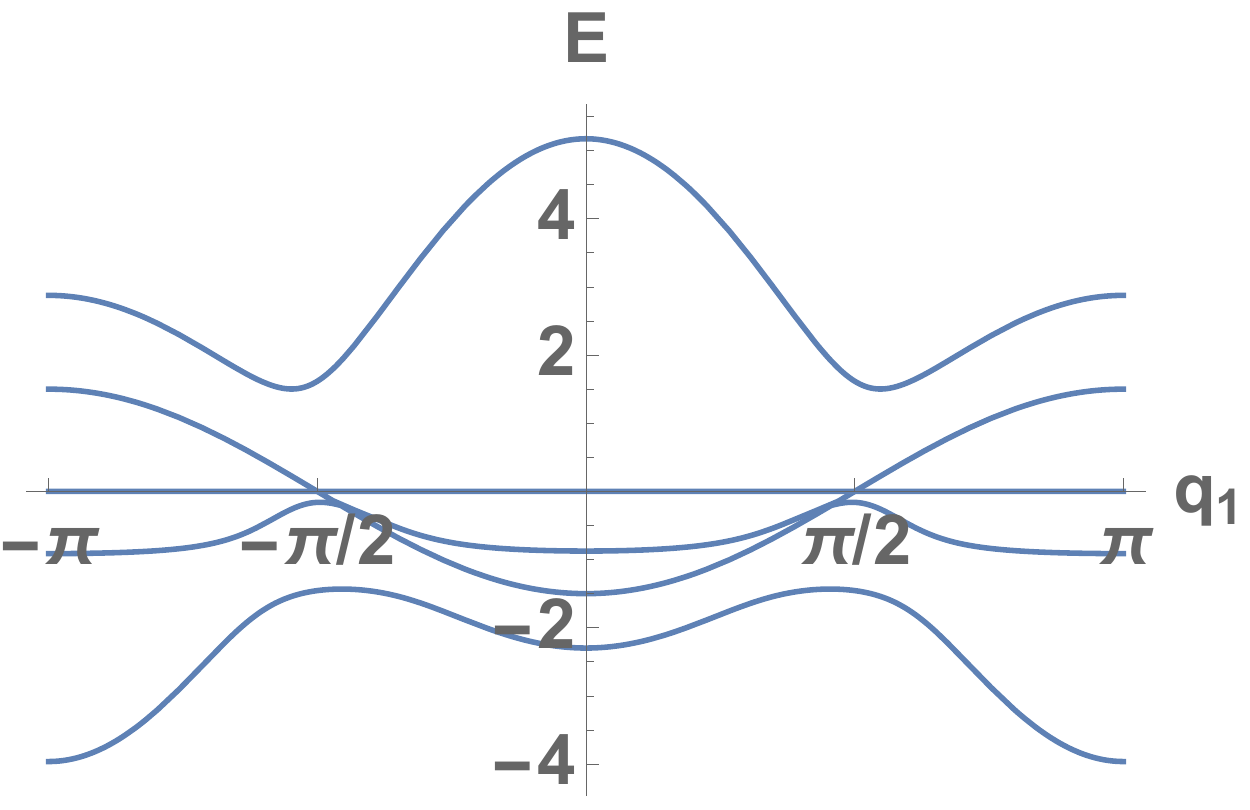}};
              \node at (\x,0) {    \includegraphics[width=0.23\textwidth,angle=0]{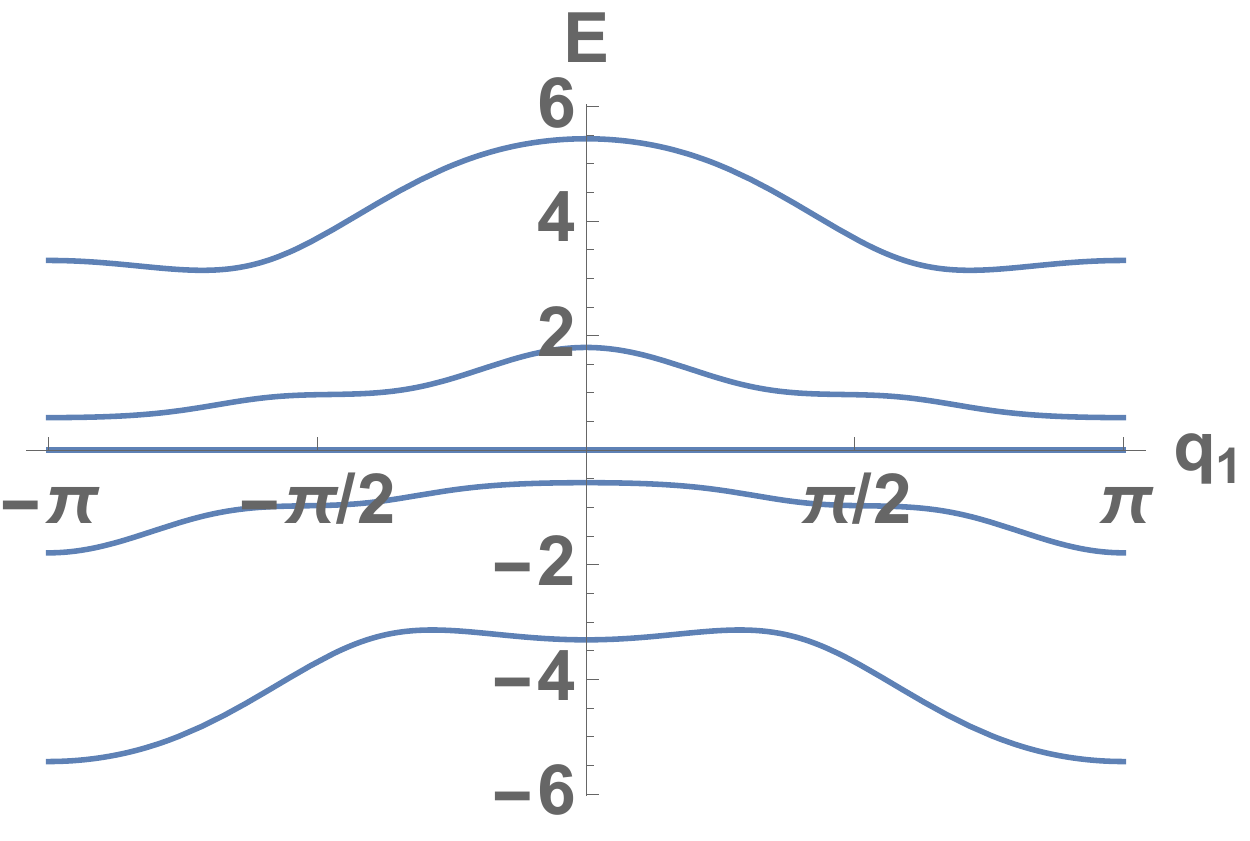}};
			 \node at (0,\y) {    \includegraphics[width=0.23\textwidth,angle=0]{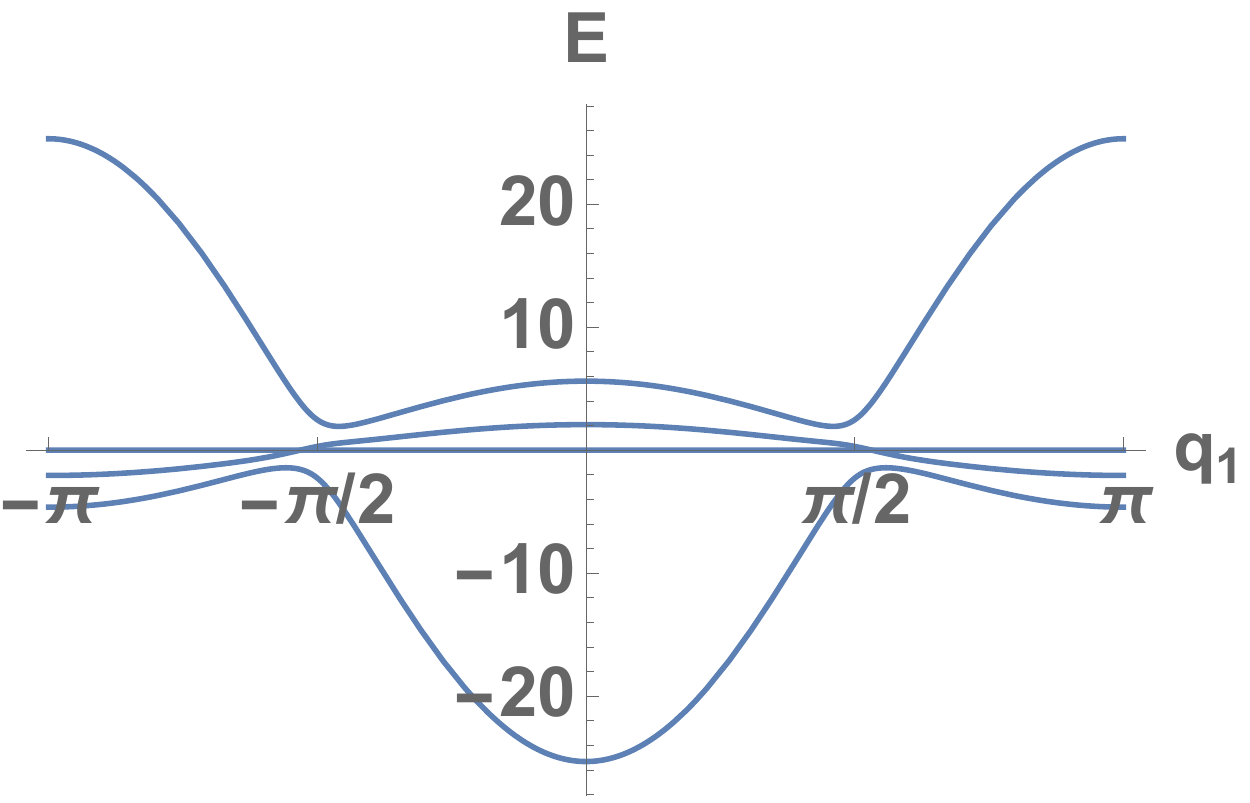}};
              \node at (\x,\y) {    \includegraphics[width=0.23\textwidth,angle=0]{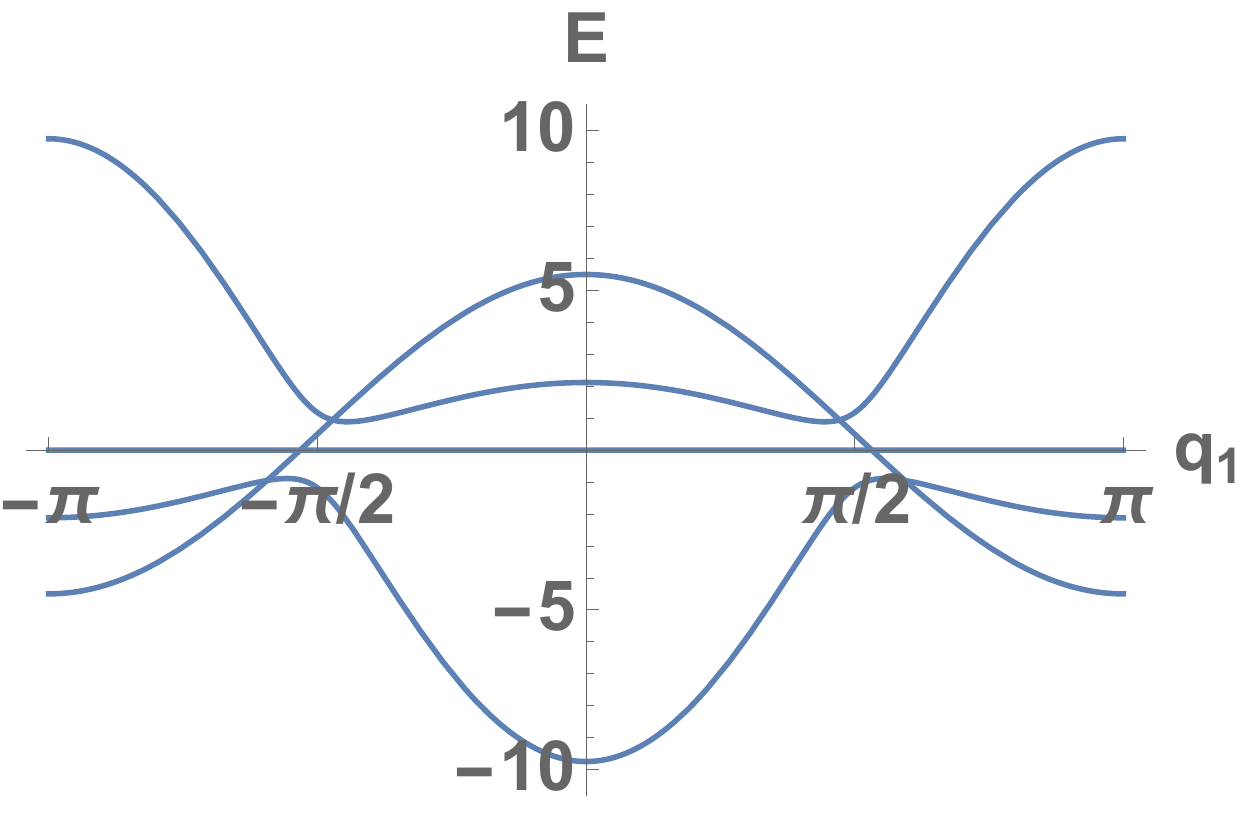}};
              \node at (0.5*\x,\y-7.0) {    \includegraphics[width=0.23\textwidth,angle=0]{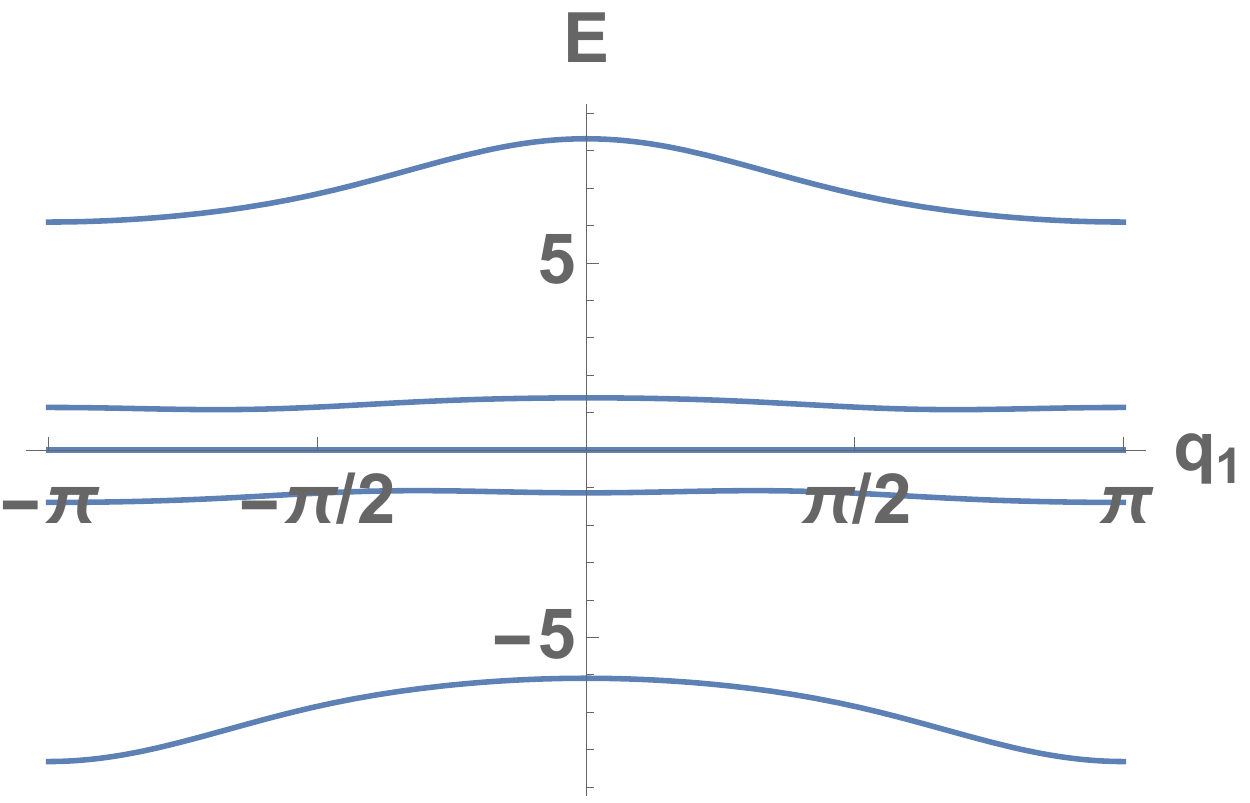}};
      \node at (-1.5,5) {(a)};
            \node at (-1.5+\x,5) {(b)};
                  \node at (-1.5,5-\y) {(c)};
                        \node at (-1.5+\x,5-\y) {(d)};
                                 \node at (-1.5+0.5*\x,-5.5-\y+7.0) {(e)};
      \end{tikzpicture}

      \caption{\label{fig:4}\textit{$(4, 1)$- and $(5,1)$-flat band lattices}. 
      (a) Eigenvalues of the Hamiltonian~\eqref{eqn:ex41a} with $r_1 = 1.0$, $r_2 =\sqrt{\pi}$, $r_3 = \sqrt{2}$, $r_4 = -1/2$, $r_5 = 2.5$, $r_6 =1/4$; 
      (b) Eigenvalues of the Hamiltonian~\eqref{eqn:ex41b} with $r_1 = 1.0$, $r_2 =\sqrt{\pi}$, $r_3 = -0.5$, $r_4 = 1/\sqrt{2}$, $r_5 = 2.5$, $r_6 = 1/4$; 
      (c) Eigenvalues of the Hamiltonian~\eqref{eqn:ex51a} with $r_1 = \sqrt{2}$, $r_2 =1/2$, $r_3 = -3/4$, $r_4 = \sqrt{\pi}$, $r_5 = 1.0$, $r_6 = -1.0$, $r_7 = 1.0$, $r_8 = -1.0$; 
      (d) Eigenvalues of the Hamiltonian~\eqref{eqn:ex51b} with $r_1 = 1/2$, $r_2 =-3/4$, $r_3 =  \sqrt{\pi}$, $r_4 =-1.0$, $r_5 = 1.5$, $r_6 = -1.5$, $r_7 = 1.0$, $r_8 = \rme$;
      (e) Eigenvalues of the Hamiltonian~\eqref{eqn:ex51e} with $r_1 = 1/2$, $r_2 =-3/4$, $r_3 =  \sqrt{\pi}$, $r_4 =-1.0$, $r_5 = 1.0$, $r_6 = -1.5$, $r_7 =  \rme$, $r_8 =1.0$.
      }
\end{figure}
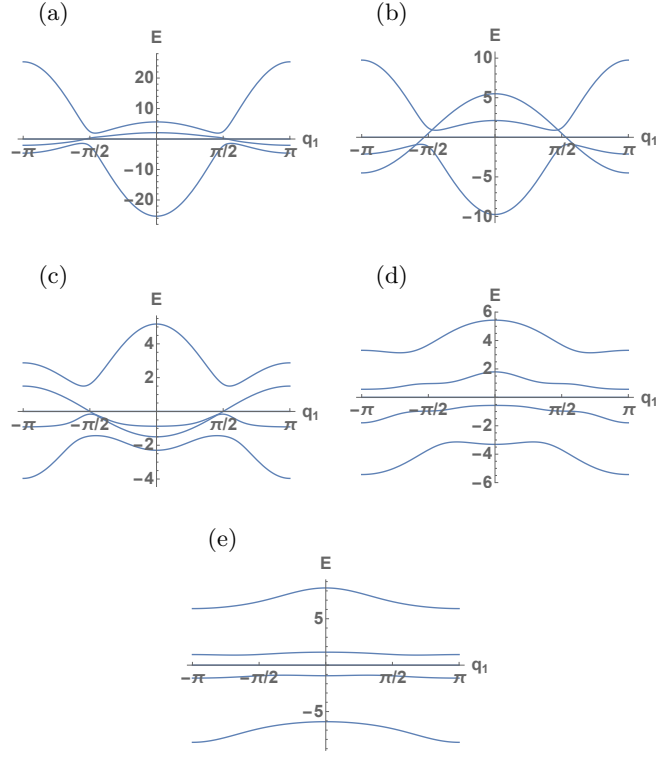

We present here a collection of one-dimensional Bloch Hamiltonians with four and five bands; that is,  we take $N = 4,5$, $M = 1$ (Fig.~\ref{fig:4}). For example, we find the following flat band Hamiltonians:
\begin{subequations}
    \begin{align}
    \label{eqn:ex41a}
        \mhq &=  \rme^{-\rmi q_1}  \begin{pmatrix}
            r_1 & 0 & 0 & 0  \\
            0 & -\frac{r_1 r_2^2}{r_4^2} & r_2 & r_3  \\
            r_4& 0 & 0 &  0 \\
            \frac{r_3 r_4}{r_2} & 0  & 0 & r_5    
        \end{pmatrix} + \begin{pmatrix}
            0 & 0 & 0 & 0 \\
            0 & 0 & 0 & 0 \\
            0 & 0 & 0 & 0 \\
            0 & 0 & 0 & r_6 
        \end{pmatrix}  + \mathrm{h.c.},\\
    \label{eqn:ex41b}
        \mhq &=  \rme^{-\rmi q_1}  \begin{pmatrix}
            r_1 & 0 & 0 & 0   \\
            0 & -\frac{r_1 (r_2 - r_4)^2}{r_3^2}   & r_2  & 0   \\
            r_3 & r_4  & -\frac{r_2 r_3^2 r_4}{r_1 (r_2 - r_4)^2}  & 0  \\
            0 & 0  & 0 & r_5     
        \end{pmatrix} \\
        \notag & + \begin{pmatrix}
            0 & 0 & 0 & 0 \\
0 & 0 & 0 & 0 \\
0 & 0 & 0 & 0 \\
0 & 0 & 0 & r_6 
\end{pmatrix}  + \mathrm{h.c.},\\
\label{eqn:ex41c}
\mathcal{H}_q &=  \rme^{-\rmi q_1}  \begin{pmatrix}
r_1  & 0 & 0 &  r_2 \\
0 & -\frac{r_1 r_4^2}{r_3^2}  & 0 & \frac{r_2 r_4}{r_3}  \\
r_3 & r_4 & 0  & 0  \\
0 & 0  & 0 & r_5     
\end{pmatrix} + \begin{pmatrix}
0 & 0 & 0 & 0 \\
0 & 0 & 0 & 0 \\
0 & 0 & 0 & 0 \\
0 & 0 & 0 & r_6 
\end{pmatrix}  + \mathrm{h.c.},\\
\label{eqn:ex51a}
\mathcal{H}_q &=  \rme^{-\rmi q_1}  \begin{pmatrix}
 0 & 0 & 0 & r_1 & 0 \\
0 & 0& 0 & r_2 & 0 \\
 0 & 0 & r_3 & 0 & 0 \\
   r_4 & 0 & 0 & r_5 & 0 \\
    0 & 0 & 0 & r_6 & 0 
\end{pmatrix} + \begin{pmatrix}
0 &0 &0 &0 &0 \\
0 &0 &0 &0 &0 \\
0 &0 &0 &0 &0 \\
0 &0 &0 &r_7 &0 \\
0 &0 &0 &0 & r_8 \\
\end{pmatrix}  + \mathrm{h.c.},\\
\label{eqn:ex51b}
\mathcal{H}_q &=  \rme^{-\rmi q_1}  \begin{pmatrix}
-\frac{r_8 r_4^2 r_5^2 - r_3 r_4^2 r_5 r_7}{r_4^2 r_7^2} & r_1 & 0 & 0 & 0 \\
0 &0 & 0& 0 & r_2 \\
r_3  & 0 & r_8 & 0 & 0 \\
  0 & r_4 &0 &  0 & 0\\
   r_5 & r_6 & r_7 & 0 &0 \\
\end{pmatrix}  + \mathrm{h.c.},\\
\label{eqn:ex51c}
\mathcal{H}_q &=  \\
\notag & \rme^{-\rmi q_1}  \begin{pmatrix}
-\frac{r_1 r_7 (2 r_3 (r_2 +r_6) + r_1 r_7)}{4 r_4 (r_2 +r_6)^2} & r_1 & 0 & 0 & 0 \\
0 &0 & 0& 0 & r_2 \\
r_3 & 0 & r_4 & 0 & 0 \\
  0 & r_5 &0 &  0 & -\frac{r_5 r_7^2}{4 r_2 r_4 + 4 r_4 r_6} \\
 \frac{r_7 (2 r_3 (r_2 + r_6) -r_1 r_7)}{4 r_4 (r_2+ r_6)}   & r_6 & r_7 & 0 &0 \\
\end{pmatrix}  \\
\notag &+ \mathrm{h.c.},\\
\label{eqn:ex51d}
\mathcal{H}_q &=  \rme^{-\rmi q_1}  \begin{pmatrix}
r_1 & r_2 & 0 & 0 & 0 \\
0 &0 & 0& 0 & r_3 \\
 r_4 & 0 & 0 & 0 & 0 \\
  0 & 0 &0 &  0 &r_5 \\
 r_6   & 0 & r_7 & 0 &0 \\
\end{pmatrix}  + \mathrm{h.c.},\\
\label{eqn:ex51e}
\mathcal{H}_q &=  \rme^{-\rmi q_1}  \begin{pmatrix}
\frac{r_1}{r_2 + r_6} \left(g_1 - \frac{r_3 r_7 }{r_8}\right) & r_1 & 0 & 0 & 0 \\
0 &0 & 0& 0 & r_2 \\
 r_3 & 0 &r_8 & 0 & 0 \\
  0 & r_4 &0 &  0 & r_5 \\
g_1   & r_6 & r_7 & 0 &0 \\
\end{pmatrix}  + \mathrm{h.c.},\\
\notag g_1 &= \frac{-2 r_2 r_8 r_5 (r_2 +r_6) - r_1 r_4 r_7^2 +r_4 (r_2 +r_6)r_3 r_7  + g_{11} }{2 r_8 r_4 (r_2 + 
   r_6)}, \\
\notag g_{11} &= -r_4 r_8(r_2 +r_6) \\
   \notag & \times     \sqrt{
\frac{(r_3 (r_2 + r_6) + r_1 r_7)^2 (4 r_8 r_5 (r_2 +r_6) + 
        r_4 r_7^2)}{r_8^2 r_4 (r_2 +r_6)^2}              
        } , \\
%
\label{eqn:ex51f}
\mathcal{H}_q &=  \rme^{-\rmi q_1}  \begin{pmatrix}
-\frac{r_1^2 r_5}{r_4 (r_2 +  r_6)} & r_1 & 0 & 0 & 0 \\
0 &0 & 0& 0 &r_2 \\
 0 & 0 & r_3& 0 & 0 \\
  0 & r_4 &0 &  0 & r_5 \\
 -\frac{r_1 r_5}{r_4}   & r_6 & 0 & 0 &0 \\
\end{pmatrix}  + \mathrm{h.c.},\\
\end{align}
\end{subequations}
such that $r_{1-8} \in \mathbb{R}$.

\bibliographystyle{apsrev4-1}
\bibliography{flatband,general,references,convex} 

\begin{thebibliography}{25}%
\makeatletter
\providecommand \@ifxundefined [1]{%
 \@ifx{#1\undefined}
}%
\providecommand \@ifnum [1]{%
 \ifnum #1\expandafter \@firstoftwo
 \else \expandafter \@secondoftwo
 \fi
}%
\providecommand \@ifx [1]{%
 \ifx #1\expandafter \@firstoftwo
 \else \expandafter \@secondoftwo
 \fi
}%
\providecommand \natexlab [1]{#1}%
\providecommand \enquote  [1]{``#1''}%
\providecommand \bibnamefont  [1]{#1}%
\providecommand \bibfnamefont [1]{#1}%
\providecommand \citenamefont [1]{#1}%
\providecommand \href@noop [0]{\@secondoftwo}%
\providecommand \href [0]{\begingroup \@sanitize@url \@href}%
\providecommand \@href[1]{\@@startlink{#1}\@@href}%
\providecommand \@@href[1]{\endgroup#1\@@endlink}%
\providecommand \@sanitize@url [0]{\catcode `\\12\catcode `\$12\catcode
  `\&12\catcode `\#12\catcode `\^12\catcode `\_12\catcode `\%12\relax}%
\providecommand \@@startlink[1]{}%
\providecommand \@@endlink[0]{}%
\providecommand \url  [0]{\begingroup\@sanitize@url \@url }%
\providecommand \@url [1]{\endgroup\@href {#1}{\urlprefix }}%
\providecommand \urlprefix  [0]{URL }%
\providecommand \Eprint [0]{\href }%
\providecommand \doibase [0]{http://dx.doi.org/}%
\providecommand \selectlanguage [0]{\@gobble}%
\providecommand \bibinfo  [0]{\@secondoftwo}%
\providecommand \bibfield  [0]{\@secondoftwo}%
\providecommand \translation [1]{[#1]}%
\providecommand \BibitemOpen [0]{}%
\providecommand \bibitemStop [0]{}%
\providecommand \bibitemNoStop [0]{.\EOS\space}%
\providecommand \EOS [0]{\spacefactor3000\relax}%
\providecommand \BibitemShut  [1]{\csname bibitem#1\endcsname}%
\let\auto@bib@innerbib\@empty
\bibitem [{\citenamefont {Chen}\ \emph {et~al.}(2005)\citenamefont {Chen},
  \citenamefont {Stajic}, \citenamefont {Tan},\ and\ \citenamefont
  {Levin}}]{Chen20051}%
  \BibitemOpen
  \bibfield  {author} {\bibinfo {author} {\bibfnamefont {Q.}~\bibnamefont
  {Chen}}, \bibinfo {author} {\bibfnamefont {J.}~\bibnamefont {Stajic}},
  \bibinfo {author} {\bibfnamefont {S.}~\bibnamefont {Tan}}, \ and\ \bibinfo
  {author} {\bibfnamefont {K.}~\bibnamefont {Levin}},\ }\href {\doibase
  http://dx.doi.org/10.1016/j.physrep.2005.02.005} {\bibfield  {journal}
  {\bibinfo  {journal} {Phys. Rep.}\ }\textbf {\bibinfo {volume} {412}},\
  \bibinfo {pages} {1 } (\bibinfo {year} {2005})}\BibitemShut {NoStop}%
\bibitem [{\citenamefont {Zwerger}(2012)}]{Zwerger2012}%
  \BibitemOpen
  \bibinfo {editor} {\bibfnamefont {W.}~\bibnamefont {Zwerger}},\ ed.,\ \href
  {\doibase 10.1007/978-3-642-21978-8} {\emph {\bibinfo {title} {{The BCS--BEC
  Crossover and the Unitary Fermi Gas}}}},\ \bibinfo {series} {Lect. Notes
  Phys.}, Vol.\ \bibinfo {volume} {836}\ (\bibinfo  {publisher} {Springer
  Berlin},\ \bibinfo {year} {2012})\BibitemShut {NoStop}%
\bibitem [{\citenamefont {Tsvelik}(2012)}]{tsvelik2012new}%
  \BibitemOpen
  \bibfield  {author} {\bibinfo {author} {\bibfnamefont {A.}~\bibnamefont
  {Tsvelik}},\ }\href@noop {} {\emph {\bibinfo {title} {New Theoretical
  Approaches to Strongly Correlated Systems}}},\ Nato Science Series II:\
  (\bibinfo  {publisher} {Springer Netherlands},\ \bibinfo {year}
  {2012})\BibitemShut {NoStop}%
\bibitem [{\citenamefont {Adams}\ \emph {et~al.}(2012)\citenamefont {Adams},
  \citenamefont {Carr}, \citenamefont {Sch\"{a}fer}, \citenamefont
  {Steinberg},\ and\ \citenamefont {Thomas}}]{1367-2630-14-11-115009}%
  \BibitemOpen
  \bibfield  {author} {\bibinfo {author} {\bibfnamefont {A.}~\bibnamefont
  {Adams}}, \bibinfo {author} {\bibfnamefont {L.~D.}\ \bibnamefont {Carr}},
  \bibinfo {author} {\bibfnamefont {T.}~\bibnamefont {Sch\"{a}fer}}, \bibinfo
  {author} {\bibfnamefont {P.}~\bibnamefont {Steinberg}}, \ and\ \bibinfo
  {author} {\bibfnamefont {J.~E.}\ \bibnamefont {Thomas}},\ }\href {\doibase
  http://dx.doi.org/10.1088/1367-2630/14/11/115009} {\bibfield  {journal}
  {\bibinfo  {journal} {New J. Phys.}\ }\textbf {\bibinfo {volume} {14}},\
  \bibinfo {pages} {115009} (\bibinfo {year} {2012})}\BibitemShut {NoStop}%
\bibitem [{\citenamefont {Laughlin}(1983)}]{PhysRevLett.50.1395}%
  \BibitemOpen
  \bibfield  {author} {\bibinfo {author} {\bibfnamefont {R.~B.}\ \bibnamefont
  {Laughlin}},\ }\href {\doibase 10.1103/PhysRevLett.50.1395} {\bibfield
  {journal} {\bibinfo  {journal} {Phys. Rev. Lett.}\ }\textbf {\bibinfo
  {volume} {50}},\ \bibinfo {pages} {1395} (\bibinfo {year}
  {1983})}\BibitemShut {NoStop}%
\bibitem [{\citenamefont {Sheng}\ \emph {et~al.}(2011)\citenamefont {Sheng},
  \citenamefont {Gu}, \citenamefont {Sun},\ and\ \citenamefont
  {Sheng}}]{Sheng11}%
  \BibitemOpen
  \bibfield  {author} {\bibinfo {author} {\bibfnamefont {D.~N.}\ \bibnamefont
  {Sheng}}, \bibinfo {author} {\bibfnamefont {Z.-C.}\ \bibnamefont {Gu}},
  \bibinfo {author} {\bibfnamefont {K.}~\bibnamefont {Sun}}, \ and\ \bibinfo
  {author} {\bibfnamefont {L.}~\bibnamefont {Sheng}},\ }\href {\doibase
  10.1038/ncomms1380} {\bibfield  {journal} {\bibinfo  {journal} {Nat.
  Commun.}\ }\textbf {\bibinfo {volume} {2}},\ \bibinfo {pages} {389} (\bibinfo
  {year} {2011})}\BibitemShut {NoStop}%
\bibitem [{\citenamefont {Neupert}\ \emph {et~al.}(2011)\citenamefont
  {Neupert}, \citenamefont {Santos}, \citenamefont {Chamon},\ and\
  \citenamefont {Mudry}}]{PhysRevLett.106.236804}%
  \BibitemOpen
  \bibfield  {author} {\bibinfo {author} {\bibfnamefont {T.}~\bibnamefont
  {Neupert}}, \bibinfo {author} {\bibfnamefont {L.}~\bibnamefont {Santos}},
  \bibinfo {author} {\bibfnamefont {C.}~\bibnamefont {Chamon}}, \ and\ \bibinfo
  {author} {\bibfnamefont {C.}~\bibnamefont {Mudry}},\ }\href {\doibase
  10.1103/PhysRevLett.106.236804} {\bibfield  {journal} {\bibinfo  {journal}
  {Phys. Rev. Lett.}\ }\textbf {\bibinfo {volume} {106}},\ \bibinfo {pages}
  {236804} (\bibinfo {year} {2011})}\BibitemShut {NoStop}%
\bibitem [{\citenamefont {Regnault}\ and\ \citenamefont
  {Bernevig}(2011)}]{PhysRevX.1.021014}%
  \BibitemOpen
  \bibfield  {author} {\bibinfo {author} {\bibfnamefont {N.}~\bibnamefont
  {Regnault}}\ and\ \bibinfo {author} {\bibfnamefont {B.~A.}\ \bibnamefont
  {Bernevig}},\ }\href {\doibase 10.1103/PhysRevX.1.021014} {\bibfield
  {journal} {\bibinfo  {journal} {Phys. Rev. X}\ }\textbf {\bibinfo {volume}
  {1}},\ \bibinfo {pages} {021014} (\bibinfo {year} {2011})}\BibitemShut
  {NoStop}%
\bibitem [{\citenamefont {Bergholtz}\ and\ \citenamefont
  {Liu}(2013)}]{doi:10.1142/S021797921330017X}%
  \BibitemOpen
  \bibfield  {author} {\bibinfo {author} {\bibfnamefont {E.~J.}\ \bibnamefont
  {Bergholtz}}\ and\ \bibinfo {author} {\bibfnamefont {Z.}~\bibnamefont
  {Liu}},\ }\href {\doibase 10.1142/S021797921330017X} {\bibfield  {journal}
  {\bibinfo  {journal} {Int. J. Mod. Phys. B}\ }\textbf {\bibinfo {volume}
  {27}},\ \bibinfo {pages} {1330017} (\bibinfo {year} {2013})}\BibitemShut
  {NoStop}%
\bibitem [{\citenamefont {Chiu}\ \emph {et~al.}(2015)\citenamefont {Chiu},
  \citenamefont {Pikulin},\ and\ \citenamefont {Franz}}]{PhysRevB.91.165402}%
  \BibitemOpen
  \bibfield  {author} {\bibinfo {author} {\bibfnamefont {C.-K.}\ \bibnamefont
  {Chiu}}, \bibinfo {author} {\bibfnamefont {D.~I.}\ \bibnamefont {Pikulin}}, \
  and\ \bibinfo {author} {\bibfnamefont {M.}~\bibnamefont {Franz}},\ }\href
  {\doibase 10.1103/PhysRevB.91.165402} {\bibfield  {journal} {\bibinfo
  {journal} {Phys. Rev. B}\ }\textbf {\bibinfo {volume} {91}},\ \bibinfo
  {pages} {165402} (\bibinfo {year} {2015})}\BibitemShut {NoStop}%
\bibitem [{\citenamefont {Flach}\ \emph {et~al.}(2014)\citenamefont {Flach},
  \citenamefont {Leykam}, \citenamefont {Bodyfelt}, \citenamefont {Matthies},\
  and\ \citenamefont {Desyatnikov}}]{flach2014detangling}%
  \BibitemOpen
  \bibfield  {author} {\bibinfo {author} {\bibfnamefont {S.}~\bibnamefont
  {Flach}}, \bibinfo {author} {\bibfnamefont {D.}~\bibnamefont {Leykam}},
  \bibinfo {author} {\bibfnamefont {J.~D.}\ \bibnamefont {Bodyfelt}}, \bibinfo
  {author} {\bibfnamefont {P.}~\bibnamefont {Matthies}}, \ and\ \bibinfo
  {author} {\bibfnamefont {A.~S.}\ \bibnamefont {Desyatnikov}},\ }\href
  {http://stacks.iop.org/0295-5075/105/i=3/a=30001} {\bibfield  {journal}
  {\bibinfo  {journal} {EPL (Europhysics Letters)}\ }\textbf {\bibinfo {volume}
  {105}},\ \bibinfo {pages} {30001} (\bibinfo {year} {2014})}\BibitemShut
  {NoStop}%
\bibitem [{\citenamefont {Leykam}\ \emph {et~al.}(2017)\citenamefont {Leykam},
  \citenamefont {Bodyfelt}, \citenamefont {Desyatnikov},\ and\ \citenamefont
  {Flach}}]{leykam2017localization}%
  \BibitemOpen
  \bibfield  {author} {\bibinfo {author} {\bibfnamefont {D.}~\bibnamefont
  {Leykam}}, \bibinfo {author} {\bibfnamefont {J.~D.}\ \bibnamefont
  {Bodyfelt}}, \bibinfo {author} {\bibfnamefont {A.~S.}\ \bibnamefont
  {Desyatnikov}}, \ and\ \bibinfo {author} {\bibfnamefont {S.}~\bibnamefont
  {Flach}},\ }\href {\doibase 10.1140/epjb/e2016-70551-2} {\bibfield  {journal}
  {\bibinfo  {journal} {Eur. Phys. J. B}\ }\textbf {\bibinfo {volume} {90}},\
  \bibinfo {pages} {1} (\bibinfo {year} {2017})}\BibitemShut {NoStop}%
\bibitem [{\citenamefont {Bodyfelt}\ \emph {et~al.}(2014)\citenamefont
  {Bodyfelt}, \citenamefont {Leykam}, \citenamefont {Danieli}, \citenamefont
  {Yu},\ and\ \citenamefont {Flach}}]{bodyfelt2014flatbands}%
  \BibitemOpen
  \bibfield  {author} {\bibinfo {author} {\bibfnamefont {J.~D.}\ \bibnamefont
  {Bodyfelt}}, \bibinfo {author} {\bibfnamefont {D.}~\bibnamefont {Leykam}},
  \bibinfo {author} {\bibfnamefont {C.}~\bibnamefont {Danieli}}, \bibinfo
  {author} {\bibfnamefont {X.}~\bibnamefont {Yu}}, \ and\ \bibinfo {author}
  {\bibfnamefont {S.}~\bibnamefont {Flach}},\ }\href {\doibase
  10.1103/PhysRevLett.113.236403} {\bibfield  {journal} {\bibinfo  {journal}
  {Phys. Rev. Lett.}\ }\textbf {\bibinfo {volume} {113}},\ \bibinfo {pages}
  {236403} (\bibinfo {year} {2014})}\BibitemShut {NoStop}%
\bibitem [{\citenamefont {Danieli}\ \emph {et~al.}(2015)\citenamefont
  {Danieli}, \citenamefont {Bodyfelt},\ and\ \citenamefont
  {Flach}}]{danieli2015flatband}%
  \BibitemOpen
  \bibfield  {author} {\bibinfo {author} {\bibfnamefont {C.}~\bibnamefont
  {Danieli}}, \bibinfo {author} {\bibfnamefont {J.~D.}\ \bibnamefont
  {Bodyfelt}}, \ and\ \bibinfo {author} {\bibfnamefont {S.}~\bibnamefont
  {Flach}},\ }\href {\doibase 10.1103/PhysRevB.91.235134} {\bibfield  {journal}
  {\bibinfo  {journal} {Phys. Rev. B}\ }\textbf {\bibinfo {volume} {91}},\
  \bibinfo {pages} {235134} (\bibinfo {year} {2015})}\BibitemShut {NoStop}%
\bibitem [{\citenamefont {Khomeriki}\ and\ \citenamefont
  {Flach}(2016)}]{khomeriki2016landau}%
  \BibitemOpen
  \bibfield  {author} {\bibinfo {author} {\bibfnamefont {R.}~\bibnamefont
  {Khomeriki}}\ and\ \bibinfo {author} {\bibfnamefont {S.}~\bibnamefont
  {Flach}},\ }\href {\doibase 10.1103/PhysRevLett.116.245301} {\bibfield
  {journal} {\bibinfo  {journal} {Phys. Rev. Lett.}\ }\textbf {\bibinfo
  {volume} {116}},\ \bibinfo {pages} {245301} (\bibinfo {year}
  {2016})}\BibitemShut {NoStop}%
\bibitem [{\citenamefont {Mielke}(1991)}]{mielke1991ferromagnetism}%
  \BibitemOpen
  \bibfield  {author} {\bibinfo {author} {\bibfnamefont {A.}~\bibnamefont
  {Mielke}},\ }\href {http://stacks.iop.org/0305-4470/24/i=14/a=018} {\bibfield
   {journal} {\bibinfo  {journal} {J. Phys. A: Math. Gen.}\ }\textbf {\bibinfo
  {volume} {24}},\ \bibinfo {pages} {3311} (\bibinfo {year}
  {1991})}\BibitemShut {NoStop}%
\bibitem [{\citenamefont {Tasaki}(1992)}]{tasaki1992ferromagnetism}%
  \BibitemOpen
  \bibfield  {author} {\bibinfo {author} {\bibfnamefont {H.}~\bibnamefont
  {Tasaki}},\ }\href {\doibase 10.1103/PhysRevLett.69.1608} {\bibfield
  {journal} {\bibinfo  {journal} {Phys. Rev. Lett.}\ }\textbf {\bibinfo
  {volume} {69}},\ \bibinfo {pages} {1608} (\bibinfo {year}
  {1992})}\BibitemShut {NoStop}%
\bibitem [{\citenamefont {Dias}\ and\ \citenamefont
  {Gouveia}(2015)}]{dias2015origami}%
  \BibitemOpen
  \bibfield  {author} {\bibinfo {author} {\bibfnamefont {R.~G.}\ \bibnamefont
  {Dias}}\ and\ \bibinfo {author} {\bibfnamefont {J.~D.}\ \bibnamefont
  {Gouveia}},\ }\href {http://dx.doi.org/10.1038/srep16852} {\bibfield
  {journal} {\bibinfo  {journal} {Sci. Rep.}\ }\textbf {\bibinfo {volume}
  {5}},\ \bibinfo {pages} {16852 EP } (\bibinfo {year} {2015})}\BibitemShut
  {NoStop}%
\bibitem [{\citenamefont {Morales-Inostroza}\ and\ \citenamefont
  {Vicencio}(2016)}]{morales2016simple}%
  \BibitemOpen
  \bibfield  {author} {\bibinfo {author} {\bibfnamefont {L.}~\bibnamefont
  {Morales-Inostroza}}\ and\ \bibinfo {author} {\bibfnamefont {R.~A.}\
  \bibnamefont {Vicencio}},\ }\href {\doibase 10.1103/PhysRevA.94.043831}
  {\bibfield  {journal} {\bibinfo  {journal} {Phys. Rev. A}\ }\textbf {\bibinfo
  {volume} {94}},\ \bibinfo {pages} {043831} (\bibinfo {year}
  {2016})}\BibitemShut {NoStop}%
\bibitem [{\citenamefont {Ramachandran}\ \emph {et~al.}(2017)\citenamefont
  {Ramachandran}, \citenamefont {Andreanov},\ and\ \citenamefont
  {Flach}}]{ramachandran2017chiral}%
  \BibitemOpen
  \bibfield  {author} {\bibinfo {author} {\bibfnamefont {A.}~\bibnamefont
  {Ramachandran}}, \bibinfo {author} {\bibfnamefont {A.}~\bibnamefont
  {Andreanov}}, \ and\ \bibinfo {author} {\bibfnamefont {S.}~\bibnamefont
  {Flach}},\ }\href {\doibase 10.1103/PhysRevB.96.161104} {\bibfield  {journal}
  {\bibinfo  {journal} {Phys. Rev. B}\ }\textbf {\bibinfo {volume} {96}},\
  \bibinfo {pages} {161104} (\bibinfo {year} {2017})}\BibitemShut {NoStop}%
\bibitem [{\citenamefont {R\"ontgen}\ \emph {et~al.}(2018)\citenamefont
  {R\"ontgen}, \citenamefont {Morfonios},\ and\ \citenamefont
  {Schmelcher}}]{roentgen2018compact}%
  \BibitemOpen
  \bibfield  {author} {\bibinfo {author} {\bibfnamefont {M.}~\bibnamefont
  {R\"ontgen}}, \bibinfo {author} {\bibfnamefont {C.~V.}\ \bibnamefont
  {Morfonios}}, \ and\ \bibinfo {author} {\bibfnamefont {P.}~\bibnamefont
  {Schmelcher}},\ }\href {\doibase 10.1103/PhysRevB.97.035161} {\bibfield
  {journal} {\bibinfo  {journal} {Phys. Rev. B}\ }\textbf {\bibinfo {volume}
  {97}},\ \bibinfo {pages} {035161} (\bibinfo {year} {2018})}\BibitemShut
  {NoStop}%
\bibitem [{\citenamefont {Maimaiti}\ \emph {et~al.}(2017)\citenamefont
  {Maimaiti}, \citenamefont {Andreanov}, \citenamefont {Park}, \citenamefont
  {Gendelman},\ and\ \citenamefont {Flach}}]{maimaiti2017compact}%
  \BibitemOpen
  \bibfield  {author} {\bibinfo {author} {\bibfnamefont {W.}~\bibnamefont
  {Maimaiti}}, \bibinfo {author} {\bibfnamefont {A.}~\bibnamefont {Andreanov}},
  \bibinfo {author} {\bibfnamefont {H.~C.}\ \bibnamefont {Park}}, \bibinfo
  {author} {\bibfnamefont {O.}~\bibnamefont {Gendelman}}, \ and\ \bibinfo
  {author} {\bibfnamefont {S.}~\bibnamefont {Flach}},\ }\href {\doibase
  10.1103/PhysRevB.95.115135} {\bibfield  {journal} {\bibinfo  {journal} {Phys.
  Rev. B}\ }\textbf {\bibinfo {volume} {95}},\ \bibinfo {pages} {115135}
  (\bibinfo {year} {2017})}\BibitemShut {NoStop}%
\bibitem [{\citenamefont {{Maimaiti}}\ \emph {et~al.}(2018)\citenamefont
  {{Maimaiti}}, \citenamefont {{Flach}},\ and\ \citenamefont
  {{Andreanov}}}]{maimaiti2018unpub}%
  \BibitemOpen
  \bibfield  {author} {\bibinfo {author} {\bibfnamefont {W.}~\bibnamefont
  {{Maimaiti}}}, \bibinfo {author} {\bibfnamefont {S.}~\bibnamefont {{Flach}}},
  \ and\ \bibinfo {author} {\bibfnamefont {A.}~\bibnamefont {{Andreanov}}},\
  }\href@noop {} {\bibfield  {journal} {\bibinfo  {journal} {in preparation}\ }
  (\bibinfo {year} {2018})}\BibitemShut {NoStop}%
\bibitem [{\citenamefont {{Rhim}}\ and\ \citenamefont
  {{Yang}}(2018)}]{rhim2018classification}%
  \BibitemOpen
  \bibfield  {author} {\bibinfo {author} {\bibfnamefont {J.-W.}\ \bibnamefont
  {{Rhim}}}\ and\ \bibinfo {author} {\bibfnamefont {B.-J.}\ \bibnamefont
  {{Yang}}},\ }\href@noop {} {\bibfield  {journal} {\bibinfo  {journal} {ArXiv
  e-prints}\ } (\bibinfo {year} {2018})},\ \Eprint
  {http://arxiv.org/abs/1808.05926} {arXiv:1808.05926} \BibitemShut {NoStop}%
\bibitem [{\citenamefont {Wan}\ \emph {et~al.}(2011)\citenamefont {Wan},
  \citenamefont {Turner}, \citenamefont {Vishwanath},\ and\ \citenamefont
  {Savrasov}}]{wan2011topological}%
  \BibitemOpen
  \bibfield  {author} {\bibinfo {author} {\bibfnamefont {X.}~\bibnamefont
  {Wan}}, \bibinfo {author} {\bibfnamefont {A.~M.}\ \bibnamefont {Turner}},
  \bibinfo {author} {\bibfnamefont {A.}~\bibnamefont {Vishwanath}}, \ and\
  \bibinfo {author} {\bibfnamefont {S.~Y.}\ \bibnamefont {Savrasov}},\ }\href
  {\doibase 10.1103/PhysRevB.83.205101} {\bibfield  {journal} {\bibinfo
  {journal} {Phys. Rev. B}\ }\textbf {\bibinfo {volume} {83}},\ \bibinfo
  {pages} {205101} (\bibinfo {year} {2011})}\BibitemShut {NoStop}%
\end{thebibliography}%

\end{document}